\documentclass[12pt]{article}
\pdfoutput=1

\usepackage[height=23truecm,width=16.4truecm,top=2.2truecm]{geometry}

\usepackage[skip=3pt,indent]{parskip}

\usepackage[utf8]{inputenc}
\usepackage{graphicx}
\usepackage{amsmath}
\usepackage{amssymb}
\usepackage{hyperref}
\hypersetup{colorlinks,citecolor=TokiwaIro,linkcolor=RuriIro,urlcolor=RuriIro,bookmarksopen,bookmarksnumbered,pdfstartview=FitH,linktocpage}
\usepackage{cite}
\usepackage{braket}
\usepackage{bm}
\usepackage{bbm}
\usepackage{color}
\usepackage[svgnames,table]{xcolor}
\usepackage{float}
\usepackage{hhline}
\usepackage{multirow}
\usepackage{indentfirst}
\usepackage{latexsym}
\usepackage{mathtools}
\usepackage{cancel}
\usepackage{slashed}
\usepackage{colortbl}
\usepackage[mathscr]{euscript}
\usepackage{enumitem}
\usepackage{subcaption}
\usepackage{comment}
\usepackage{amsthm}

\numberwithin{equation}{section}

\allowdisplaybreaks

\definecolor{RuriIro}{rgb}{0.,0.28,0.60}
\definecolor{TokiwaIro}{rgb}{0.,0.39,0.16}

\definecolor{dred}{rgb}{0.7,0.15,0.09}
\definecolor{dblue}{rgb}{0,0.12,0.64}
\definecolor{dgreen}{rgb}{0.2,0.51,0.19}

\newcommand{\nn}{\nonumber}

\newcommand{\mc}{\mathcal}
\newcommand{\mr}{\mathrm}

\newcommand{\mbb}{\mathbb}

\newcommand{\msc}{\mathscr}

\newcommand{\del}{\partial}
\newcommand{\ol}{\overline}

\newcommand{\sld}{\slashed}

\renewcommand{\Re}{\mathop{\mathrm{Re}}}

\newcommand{\diag}{\mathop{\mathrm{diag}}}

\newcommand{\rU}{\mathrm{U}}
\newcommand{\TRS}{T_{\mathcal{RS}}}
\newcommand{\TSS}{T_{\mathcal{SS}}}
\newcommand{\iso}{\mathrm{iso}}

\newcommand{\cPR}{\mathcal{P_R}}

\newcommand{\phis}{\phi_*}
\newcommand{\chis}{\chi_*}

\begin{document}

\begin{titlepage}

\begin{flushright}
{\ttfamily
KUNS-3009
}
\end{flushright}

\vspace{1cm}

\begin{center}

{\Large \bfseries%
Two-field inflation from one complex scalar\\[2mm] with symmetry breaking}

\vspace{1cm}

\renewcommand{\thefootnote}{\fnsymbol{footnote}}
{%
\hypersetup{linkcolor=black}
Yoshihiko Abe$^{1}$\footnote[1]{yabe3@wisc.edu},
Toshimasa Ito$^{2}$\footnote[2]{toshimasa.i@gauge.scphys.kyoto-u.ac.jp}, 
and Koichi Yoshioka$^{2}$\footnote[3]{yoshioka@gauge.scphys.kyoto-u.ac.jp}
}%
\vspace{5mm}

{\itshape%
$^1${Department of Physics, University of Wisconsin-Madison, Madison, WI 53706, USA}\\
$^2${Department of Physics, Kyoto University, Kyoto 606-8502, Japan}
}%

\vspace{10mm}

\end{center}

\abstract{
\noindent
We study two-field inflation derived from a single complex scalar 
field with a nonzero vacuum expectation value. The dynamics of inflation 
are governed by two parameters, the vacuum expectation value and the 
mass parameter of the phase mode, which together give rise to a rich variety 
of inflationary structures. We classify the possible trajectories of the 
two inflaton fields and identify the parameter regions consistent with 
current cosmological observations. Furthermore, we investigate the reheating 
process through the inflaton decay to right-handed neutrinos and the subsequent 
generation of lepton number within these regions. Our findings suggest that 
the presence of multiple scalar degrees of freedom can significantly alter 
the conditions for successful reheating and leptogenesis.
}

\end{titlepage}

\renewcommand{\thefootnote}{\arabic{footnote}}
\setcounter{footnote}{0}
\setcounter{page}{1}

\tableofcontents
\bigskip

\section{Introduction}

Inflation in the early universe provides a natural solution to the horizon 
and flatness problems of the big bang theory and generates the primordial 
perturbations~\cite{Starobinsky:1980te,Guth:1980zm,Linde:1981mu,Albrecht:1982wi,Starobinsky:1982ee,Linde:1983gd}. The inflationary expansion 
driven by a single scalar field has been extensively investigated. 
Moreover, multi-field inflation models have also been studied with field 
theoretical motivations~\cite{Berg:2009tg,McDonald:2014oza,Barenboim:2014vea,Achucarro:2015rfa,Ross:2016hyb,McDonough:2020gmn,Deshpande:2021ptg,Gong:2021zem,Lee:2022fkd,Khan:2023snv} and in the contexts of supergravity and string theory~\cite{Brustein:1992nk,Lalak:2007aj,Berglund:2009uf,Baumann:2014nda,Cicoli:2023opf}.

We consider a two-field inflation model arising from a single complex scalar 
field that has a soft-breaking mass and a non-minimal coupling to gravity, and 
develops a non-vanishing vacuum expectation value (VEV). After the breaking of 
the phase rotation symmetry, a pseudo Nambu-Goldstone boson (pNGB) appears as 
a massive state due to the soft-breaking mass. This pNGB exhibits properties 
similar to moduli fields arising from the dimensional reduction in string 
theory, where the potential becomes periodic because of the discretized shift 
symmetry. In addition, the pNGB can serve as an attractive dark matter 
candidate~\cite{Barger:2010yn,Gross:2017dan}, where its derivative couplings 
play an important role in satisfying the observational constraints. 
Introducing right-handed (RH) neutrinos with Yukawa couplings to the complex 
scalar field can explain the origin of neutrino masses through the scalar 
field dynamics, in which the pNGB corresponds to the 
majoron~\cite{Chikashige:1980qk,Chikashige:1980ui}. Thus, a framework
extended by a single complex scalar field provides a simple yet
powerful setting for understanding various features of theories beyond
the Standard Model (BSM).

Among the possible applications of such an extended scalar sector, one of the 
most interesting and significant is the inflationary expansion in the early 
universe. The additional complex scalar field introduces two real dynamical 
scalar components, the radial mode and the pNGB. Considering only one of 
these as the inflaton can reproduce the dynamics of well-known
single-field models such as chaotic inflation~\cite{Linde:1983gd},
natural inflation~\cite{Freese:1990rb,Adams:1992bn,Savage:2006tr,Stein:2021uge,Salvio:2021lka,Mukuno:2024yoa}, and Higgs inflation~\cite{Bezrukov:2007ep,Rubio:2018ogq,Enckell:2018uic,Koshelev:2020fok,Pareek:2023die}. 
A radiatively generated potential for the radial mode has also been
used in the context of the Majoron model for inflation~\cite{Boucenna:2014uma}. 
Which inflationary dynamics are realized is determined by model
parameters such as the VEV and the soft-breaking mass parameter of the pNGB. 
Most previous analyses of inflation driven by a complex scalar field
have focused on situations effectively reducible to single-field
inflation. However, when multiple scalar degrees of freedom are
dynamically involved, particularly in extensions with a complex scalar
field that can serve as a simplified model of various classes of BSM
scenarios, it becomes important to examine the full two-field dynamics
and the resulting implications for phenomenology.

In this work, we conduct a comprehensive analysis of the parameter space for 
two-field inflation from a single complex scalar with spontaneous symmetry 
breaking, to investigate its role in inflation and in the generation of the 
matter–antimatter asymmetry of the universe. The inflationary trajectories 
are classified according to which field components dominate the inflationary 
expansion, while varying the VEV and the soft-breaking mass
parameter. We then present the parameter regions of our model
consistent with current cosmological observations. This classification
of trajectories clarifies the typical features and predictions of each
inflationary regime. We find that Higgs-like inflation driven by the
radial mode is favored over a broad range of parameters, while other
types can also produce successful inflation in this two-field
framework. Furthermore, we explore the possibilities of reheating and
leptogenesis through two inflaton decays into RH neutrinos.

This paper is organized as follows. Section~\ref{sec:Inflaton} introduces the 
model and classifies the inflationary trajectories into three representative 
categories. Section~\ref{sec:constraints} presents the parameter regions 
consistent with the current cosmological observations. In 
Section~\ref{sec:reheating}, we discuss the reheating process via inflaton 
decay into RH neutrinos and evaluate the corresponding reheating temperature 
relevant for thermal leptogenesis. Finally, Section~\ref{sec:conclusions} 
summarizes our conclusions. The appendices provide detailed
definitions and notations of parameters, and supplementary analyses of
the inflaton dynamics.

\medskip

\section{Motions of inflaton with symmetry breaking}
\label{sec:Inflaton}

We first introduce our two-field inflation model and analyze its
inflationary dynamics by solving the equations of motion (EOMs).
Throughout this paper, we set the reduced Planck mass $M_P \approx 2.4\times 10^{18}~\mr{GeV}$ 
to unity unless stated otherwise. We adopt the Minkowski metric with
signature $\eta_{\mu\nu} = \diag(+1, -1, -1, -1)$.

\subsection{Complex scalar with non-minimal coupling}
\label{subsec:Model}

We consider a complex scalar field $\Phi$ with a non-minimal coupling to
gravity. Its action is given by
\begin{align}
  S=\int d^{4}x\sqrt{-g_{J}} \left[-\frac{1}{2}\Omega^{2}(\Phi)R_{J}+g^{\mu\nu}_{J}\partial_{\mu}\Phi\partial_{\nu}\Phi^{*}-U(\Phi,\Phi^{*})\right],
  \label{formula:JordanAction}
\end{align}
where we introduce the non-minimal coupling $\xi$,
\begin{align}
  \Omega^{2}(\Phi)=1+2\xi|\Phi|^{2}.
\end{align}
$R_{J}$ is the Ricci scalar constructed from the Jordan-frame metric
$g_{J\mu\nu}$. We take the potential $U(\Phi,\Phi^*)$ to be
\begin{align}
  U(\Phi,\Phi^{*})=\frac{\lambda}{2}|\Phi|^{4}-\frac{\mu_{\Phi}^{2}}{2}|\Phi|^{2}-\frac{m_{\chi}^{2}}{4}(\Phi^{2}+\Phi^{*2})+U_{0},
  \label{Formula:PotentialU}
\end{align}
as considered in Ref.~\cite{Gross:2017dan} in the context of dark
matter. The third term explicitly breaks the $\rU(1)$ phase-rotation
symmetry of $\Phi$ down to $\mbb{Z}_2$ and generates the pseudo
Nambu-Goldstone boson (pNGB) mass. The parameter $m_\chi (\geq0)$ is
referred to as the soft-breaking mass. Under the transformation $\Phi
\mapsto e^{i \pi /2} \Phi$, the action is invariant under the sign flip
$m_\chi^2 \mapsto - m_\chi^2$, which implies that we can consistently
impose $m_\chi^2 \geq 0$ in the model. The ultraviolet completeness of
this soft-breaking mass term is not specified in this paper, but
possible origins are discussed in several
works~\cite{Abe:2020iph,Okada:2020zxo,Abe:2021byq,Okada:2021qmi,Abe:2022mlc,Liu:2022evb}.

We assume that the scalar field $\Phi$ acquires a non-vanishing VEV
$v_\phi$. The stationary condition, together with the requirement that
$U=0$ in the vacuum, yields relations among the parameters $\mu_\Phi$,
$U_0$, and $v_\phi$. If we parameterize $\Phi$ in the non-linear
representation as 
\begin{align}
  \Phi = \frac{\phi}{\sqrt{2}} e^{i \chi},
\end{align}
the scalar potential for the fields $\phi$ and $\chi$ can be written as
\begin{align}
  U(\phi,\chi)=\frac{\lambda}{8}(\phi^{2}-v_{\phi}^{2})^{2}+\frac{m_{\chi}^{2}}{4}\phi^{2}(1-\cos2\chi).
  \label{formula:PotentialUtwo}
\end{align}
Note that $\chi$ is a dimensionless variable.

To move from the Jordan frame to the Einstein frame~\cite{Hirai:1993cn,Kaiser:1994vs,Bezrukov:2007ep,Kaiser:2010ps,Abedi:2014mka,Koshelev:2020fok,Pareek:2023die}, we perform the field-dependent Weyl rescaling
\begin{align}
  g_{\mu\nu} = \Omega^2 g_{J\mu\nu} = (1 + \xi \phi^2) g_{J\mu\nu},
\end{align}
where $g_{\mu\nu}$ denotes the metric in the Einstein frame. This
redefinition leads to the following action for $\phi$ and $\chi$:
\begin{align}
  S = \int d^4x \sqrt{-g} \biggl[
  - \frac{R}{2} 
  + \frac{1 + \xi \phi^2 + 6 \xi^2 \phi^2}{2(1 + \xi \phi^2)^2} \del_\mu \phi \del^\mu \phi + \frac{\phi^2}{2(1 + \xi \phi^2)} \del_\mu \chi \del^\mu \chi
  - V(\phi, \chi)
  \biggr],
  \label{formula:Action}
\end{align}
where $R$ represents the Ricci scalar of $g_{\mu\nu}$, and the inflaton
potential is given by
\begin{align}
  V(\phi, \chi) \coloneqq \frac{U(\phi, \chi)}{(1 + \xi \phi^2)^2}
  = \frac{\lambda (\phi^2 - v_\phi^2)^2 + 2m_\chi^2 \phi^2 ( 1 - \cos 2\chi)}{8(1 + \xi \phi^2)^2}.
  \label{Formula:PotentialV}
\end{align}
In the following, we refer to the first term in \eqref{Formula:PotentialV}
as the Higgs potential and to the second term as the pNGB potential,
respectively. Note that the pNGB potential differs from the usual
natural inflation potential in
Refs.~\cite{Freese:1990rb,Adams:1992bn,Savage:2006tr,Salvio:2021lka,Stein:2021uge,Mukuno:2024yoa} in that the overall coefficient of the
potential depends on $\phi$, which in our case arises naturally from the
pNGB soft-breaking term.

The field-space metric $K_{ab}(\varphi)$ for real scalar fields
$\bm{\varphi}$ is generally defined from the kinetic term of the
Lagrangian as
\begin{align}
  \bm{L}_{\text{kin.}} = \frac{1}{2} K_{ab}(\varphi) \del_\mu \varphi^a \del_\nu \varphi^b g^{\mu\nu},
\end{align}
where $\bm{\varphi}$ is the multiplet of scalar fields and the indices
$a$ and $b$ run over all components. In the present model, 
$\varphi^1 = \phi$ and $\varphi^2 = \chi$, and their metric is read off from
\eqref{formula:Action} as
\begin{align}
  K_{ab} = \diag \left(\frac{1+\xi\phi^{2}+6\xi^{2}\phi^{2}}{(1+\xi\phi^{2})^{2}},\frac{\phi^{2}}{1+\xi\phi^{2}}\right).
\end{align}

\subsection{Slow-roll and slow-turn approximation}
\label{sec:SRSTparameter}

\subsubsection{Equations of motion and inflation parameters}

The equations of motion for the scalar fields are given by 
\begin{align}
  \frac{d^{2}\phi}{dN^{2}}+\gamma^{1}_{11}\left(\frac{d\phi}{dN}\right)^{2}+\gamma^{1}_{22}\left(\frac{d\chi}{dN}\right)^{2}+(3-\varepsilon)\left(\frac{d\phi}{dN}+K^{11}\frac{\partial}{\partial\phi}\ln V\right) &=0,
  \label{formula:phiEoM} 
    \\
  \frac{d^{2}\chi}{dN^{2}}+2\gamma^{2}_{12}\frac{d\phi}{dN}\frac{d\chi}{dN}+(3-\varepsilon)\left(\frac{d\chi}{dN}+K^{22}\frac{\partial}{\partial\chi}\ln V\right) &=0,
  \label{formula:chiEoM}
\end{align}
where the number of e-folds $N$ is defined by $d N = H dt$, with $H$
being the Hubble parameter, and $\gamma^a_{bc}$ is the Levi-Civita
connection derived from the field-space metric $K_{ab}$. The slow-roll
(SR) parameter $\varepsilon$ is defined by 
\begin{align}
  \varepsilon \coloneqq - \frac{\dot{H}}{H^2} 
  = \frac{1}{2} K_{11}\left(\frac{d\phi}{dN}\right)^{2} +
  \frac{1}{2} K_{22}\left(\frac{d\chi}{dN}\right)^{2}, 
  \label{formula:nonepsilon}
\end{align}
where the dot denotes the derivative with respect to the cosmic time
$t$. For further details, see Appendix~\ref{app:scalar-sector}. In order
to realize SR inflation, we impose the first SR condition
\begin{align}
  \varepsilon \ll 1,
  \label{formula:epsilonsmall}
\end{align}
as discussed in
Refs.~\cite{Sasaki:1998ug,GrootNibbelink:2000vx,GrootNibbelink:2001qt,Seery:2005gb,Langlois:2008mn,Peterson:2010np,Gong:2016qmq,Christodoulidis:2018qdw}. This condition also appears in the case of single-field inflation.

In the present model, the acceleration of the inflaton fields is
characterized by the parameters 
\begin{align}
  \eta^{1} &= \frac{d^{2}\phi}{dN^{2}}+\gamma^{1}_{11}\left(\frac{d\phi}{dN}\right)^{2}+\gamma^{1}_{22}\left(\frac{d\chi}{dN}\right)^{2},
  \\
  \eta^{2}&=\frac{d^{2}\chi}{dN^{2}}+2\gamma^{2}_{12}\frac{d\phi}{dN}\frac{d\chi}{dN},
\end{align}
in the $\phi$- and $\chi$-directions, respectively. Since there are two
inflaton fields, it is useful to decompose the dynamics into the
direction tangential to the inflationary trajectory and the direction of
its rotation~\cite{Gordon:2000hv,GrootNibbelink:2000vx,GrootNibbelink:2001qt}. The former is called the parallel direction, and
its unit vector in field space is denoted by $\hat{\bm{e}}_\parallel$.
The latter is the direction perpendicular to $\hat{\bm{e}}_\parallel$,
and its unit vector is denoted by $\hat{\bm{e}}_\perp$. Using these unit
vectors, the conditions on the acceleration for successful inflation are
given by 
\begin{align}
  \Bigl| \frac{\eta_\parallel}{v} \Bigr| &\coloneqq \biggl| \frac{K_{ab} \hat{e}^a_\parallel \eta^b}{v} \biggr| \ll 1,
  \label{formula:nonetaparallel}
  \\
  \frac{\eta_\perp}{v} &\coloneqq \frac{K_{ab} \hat{e}^a_\perp \eta^b}{v} \ll 1,
  \label{formula:nonetaperp}
\end{align}
where $v = \sqrt{2 \varepsilon}$ is the speed of the field vector in
field space~\cite{Sasaki:1998ug,GrootNibbelink:2000vx,GrootNibbelink:2001qt,Peterson:2010np}. Equation \eqref{formula:nonetaparallel} is the second SR condition, while 
Eq.~\eqref{formula:nonetaperp} is referred to as the slow-turn (ST)
condition. The ST condition quantifies the smallness of the
perpendicular acceleration. Once the SR conditions
\eqref{formula:epsilonsmall}, \eqref{formula:nonetaparallel} and the ST
condition \eqref{formula:nonetaperp} are satisfied, the SR and ST
parameters can be approximately expressed in terms of the inflaton
potential $V$~\cite{Peterson:2010np},
\begin{align}
  \varepsilon & \approx \frac{1}{2} K^{ab} \del_a \ln V \del_b \ln V,
  \label{formula:epsilon}
  \\
  \frac{\eta_\parallel}{v} &\approx - \hat{e}_\parallel^a 
  (\del_a \del_b \ln V-\gamma^c_{ab}\del_c\ln V)\hat{e}^b_\parallel,
  \label{formula:etaPov}
  \\
  \frac{\eta_\perp}{v} &\approx - \hat{e}_\parallel^a 
  (\del_a \del_b \ln V-\gamma^c_{ab}\del_c\ln V)\hat{e}^b_\perp.
  \label{formula:etaPerpov}
\end{align}
The unit vectors in the parallel and perpendicular directions are
written as
\begin{align}
  \hat{e}_{\parallel}^a &= \frac{1}{v}\,\frac{d\varphi^a}{dN}
  \approx \frac{-K^{ab}\del_b \ln V}{\sqrt{K^{cd} \del_c \ln V \del_d \ln V}}, 
  \label{formula:eparallel}  \\
  \hat{e}_\perp^a &\approx \frac{\pm(K^{ab}-\hat{e}_\parallel^a\hat{e}_\parallel^b)
  (\del_b \del_c \ln V-\gamma^d_{bc}\del_d\ln V)\hat{e}_\parallel^c}{\sqrt{(K^{ab}-\hat{e}_\parallel^a\hat{e}_\parallel^b)
  (\del_a \del_c \ln V-\gamma^e_{ac}\del_e\ln V)
  (\del_b \del_d \ln V-\gamma^f_{bd}\del_f\ln V)\hat{e}_\parallel^c \hat{e}_\parallel^d}}.
  \label{formula:eperp}
\end{align}
The direction of $\hat{\bm{e}}_\perp$ is chosen such that
$\eta_\perp\geq 0$. By using the EOMs, one can show that the above
expression for the perpendicular unit vector is equivalent to
$K^{ab}J_{bc}\hat{e}_\parallel^c$ with some antisymmetric $J_{ab}$,
which is another representation consistent with the definition of being
perpendicular to $\hat{e}_\parallel$. In the case of a two-dimensional
field space, a simple choice is $J_{ab}=\sqrt{\det K}/v\epsilon_{ab}$,
where $\epsilon_{ab}$ is the totally antisymmetric tensor. Note that all
of the above quantities are written in terms of the fields and their
derivatives through the potential $V$ and the field-space metric
$K_{ab}$, whose values are obtained by solving the equations of motion
\eqref{formula:phiEoM} and \eqref{formula:chiEoM}.

\subsection{Classification of inflation trajectory}
\label{sec:classification}

\begin{figure}[t]
  \centering
  \includegraphics[width=0.48\textwidth]{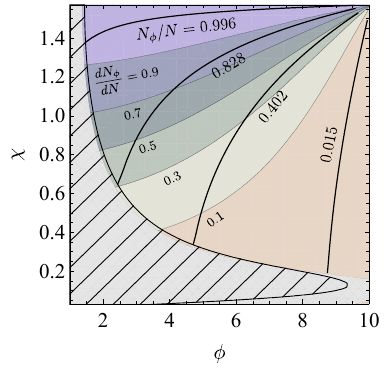}
  \quad
  \includegraphics[width=0.47\textwidth]{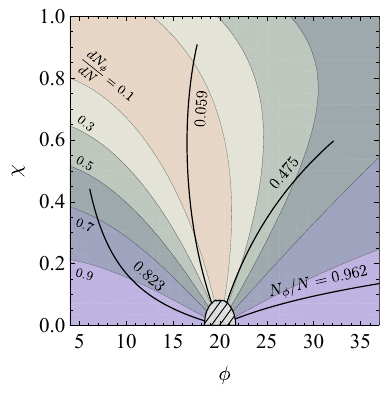}
  \caption{
    Typical inflation trajectories in the present two-field model with
    various e-folding ratios $N_\phi /N$. The model parameters are
    chosen as $v_\phi =10^{-4}$, $m_\chi = 10^{-4.5}$, $\xi=10^{-2}$,
    $\lambda = 10^{-12.6}$ (Left), and $v_\phi = 20$, $m_\chi = 10^{-6}$,
    $\xi = 10^{-3}$, $\lambda=10^{-14.6}$ (Right). The total e-folding
    number is fixed to $N=60$. The color density shows the value of
    $d N_\phi / dN$, which is the integrand in \eqref{eq:def-N_phi}. The
    gray and shaded regions are excluded by the inflationary conditions
    $\varepsilon < 1$ and $\eta_\parallel < 1$, respectively.
  }
  \label{fig:Nphi-trajectory}
  \bigskip
\end{figure}

In this subsection, we classify the inflationary trajectories in the
present two-field inflaton model. This classification allows us to
understand the multi-field nature of inflation without examining in
detail the complicated dynamics of the scalar fields arising from their
mutual interactions.

To quantify how the $\phi$ component contributes to the inflationary
expansion, we define 
\begin{align}
  N_\phi = \int_0^{N} dN' \frac{K_{11} \left( \frac{\del \phi}{\del N'} \right)^2}{K_{11} \left( \frac{\del \phi}{\del N'} \right)^2 + K_{22} \left( \frac{\del \chi}{\del N'} \right)^2},
  \label{eq:def-N_phi}
\end{align}
which can be interpreted as the effective number of e-folds generated by
the evolution of $\phi$. The integrand $dN_\phi/dN$ represents the local
fraction of e-folds in the $\phi$ direction and is determined from the
gradient of the scalar potential. The total ratio $N_\phi / N$ serves as
an indicator distinguishing which field component mainly drives the
inflationary expansion. Figure~\ref{fig:Nphi-trajectory} shows typical
inflation trajectories in the present two-field model together with
their e-folding ratios $N_\phi / N$. We adopt the parameter values
$v_\phi =10^{-4}$, $m_\chi = 10^{-4.5}$, $\xi=10^{-2}$, and $\lambda =
10^{-12.6}$ in the left panel, and $v_\phi = 20$, $m_\chi = 10^{-6}$,
$\xi = 10^{-3}$, and $\lambda=10^{-14.6}$ in the right panel. The
e-folding number is fixed to $N=60$ in both cases. The gray and shaded
regions are excluded by the conditions $\varepsilon < 1$ and
$\eta_\parallel < 1$, respectively. As seen in
Figure~\ref{fig:Nphi-trajectory}, the inflaton fields roll predominantly
along one scalar-field direction for large or small values of
$N_\phi /N$, while both directions contribute to inflation for
intermediate values of $N_\phi/N$.

Based on this behavior, we classify the inflationary trajectories in the
present two-field model into the following three categories:
\begin{itemize}
\item Higgs-inflation type driven mainly by $\phi$ : 
$N_\phi /N > 0.9$
\item Mixed type with both $\phi$ and $\chi$ acting as inflatons : 
$0.1 \leq N_\phi/N \leq 0.9$	
\item Natural-inflation type driven mainly by $\chi$ : 
$N_\phi/N <0.1$
\end{itemize}
In the following section, we examine the relationship between the
inflationary trajectories and the corresponding cosmological observables.
We use the terms ``Higgs'', ``Mixed'', and ``Natural'' to specify which type
of inflationary trajectory we refer to.

\medskip

\section{Constraints from cosmological observations}
\label{sec:constraints}

This section presents the analysis of the inflationary observables and
the corresponding parameter constraints for the present two-field
inflation model.

\subsection{Transfer functions and observables}

In single-field inflation, the observables are derived from the
fluctuations of the inflaton field and the curvature perturbation,
denoted by $\mc{R}$, in the comoving gauge on hypersurfaces of constant
time~\cite{Bardeen:1980kt,Kodama:1984ziu,Lyth:1984gv,Sasaki:1986hm}. In
multi-field inflation, this scalar fluctuation is defined as the
projection of the multi-field perturbations onto the tangent direction
of the inflationary trajectory:
\begin{align}
  \mc{R} = \frac{K_{ab}\hat{e}_\parallel^a \delta \varphi_f^b}{v}.
  \label{Formula:R}
\end{align}
The perturbation $\delta \bm{\varphi}_f$ is the multi-field extension of
the Mukhanov–Sasaki variable~\cite{Sasaki:1986hm,Mukhanov:1990me,Peterson:2010np}:
\begin{align}
  \delta\bm{\varphi}_{f} = \delta\bm{\varphi} + \psi\frac{d\bm{\varphi}}{d N},
\end{align}
where $\delta\bm{\varphi}$ denotes the scalar-field fluctuation and
$\psi$ is a component of the diagonal metric
perturbation~\cite{Bardeen:1980kt,Kodama:1984ziu,Mukhanov:1990me}.
Furthermore, one must consider the fluctuations in the direction
orthogonal to the inflationary trajectory, which give rise to the
isocurvature perturbation~\cite{Wands:2002bn,Peterson:2010np}:
\begin{align}
  \mc{S} = \frac{K_{ab}\hat{e}_\perp^a \delta \varphi_f^b}{v}.
  \label{Formula:S}
\end{align}
These two perturbations, $\mc{R}$ and $\mc{S}$, are related by the
transfer functions~\cite{Amendola:2001ni,Wands:2002bn,Peterson:2010np,Karamitsos:2017elm}:
\begin{align}
  \begin{pmatrix}
    \mc{R} \\
    \mc{S}
  \end{pmatrix}
  = \begin{pmatrix}
      1 & \TRS \\
      0 & \TSS
    \end{pmatrix}
  \begin{pmatrix}
    \mc{R}_* \\
    \mc{S}_*
  \end{pmatrix}.
  \label{Formula:RST}
\end{align}
In this paper, quantities evaluated at the horizon exit are indicated by
an asterisk, as in $\mc{R}_*$ and $\mc{S}_*$. This relation shows that
the curvature perturbation $\mc{R}$ is not frozen after the horizon exit,
and the isocurvature perturbation $\mc{S}$ contributes to it through the
transfer function $\TRS$~\cite{Lyth:1984gv,GrootNibbelink:2000vx,GrootNibbelink:2001qt,DiMarco:2002eb,Karamitsos:2017elm,Enckell:2018uic}.

In contrast to single-field inflation, $\TRS$ can significantly affect
the inflationary observables in the multi-field case. The power spectrum
$\cPR$ and tensor-to-scalar ratio $r$ are given
by~\cite{Sasaki:1998ug,Peterson:2010np}
\begin{align}
  \cPR = \frac{V_*}{24\pi^2 \varepsilon_*} ( 1 + \TRS^2),
  \qquad 
  r = \frac{16 \varepsilon_*}{1 + \TRS^2}.
\end{align}
In addition, we define the spectral index $n_s$ and the running
spectrum $\alpha$ as
\begin{align}
  n_s = 1 + \frac{d \ln \cPR}{d \ln k},
  \qquad 
  \alpha = \frac{d n_s}{d \ln k},
\end{align}
and the isocurvature fraction~\cite{Beltran:2005gr,Hamann:2009yf,Karamitsos:2017elm,Enckell:2018uic}:
\begin{align}
  \beta_\iso = \frac{\TSS^2}{1 + \TRS^2 + \TSS^2}.
  \label{formula:betaiso}
\end{align}
These quantities depend on the transfer
functions~\cite{Sasaki:1995aw,Peterson:2010np,Enckell:2018uic,Karamitsos:2017elm},
and all of them can be expressed in terms of the parameters and field
values in the present inflationary model. For details, see
Appendix~\ref{app:observables}. 

These inflationary observables are experimentally constrained by the
Planck Collaboration~\cite{Planck:2018jri}:
\begin{gather}
  \cPR = (2.1 \pm 0.1) \times 10^{-9},
  \qquad 
  r< 0.07,  \qquad
  n_s = 0.9649 \pm 0.0042,  \nonumber \\
  \alpha = -0.0045 \pm 0.0067, \qquad 
  \beta_\iso < 0.038.
  \label{eq:obs-constraints}
\end{gather}

\subsection{Spectral index and running spectrum}

\begin{figure}[t]
  \centering
  \includegraphics[width=0.48\textwidth]{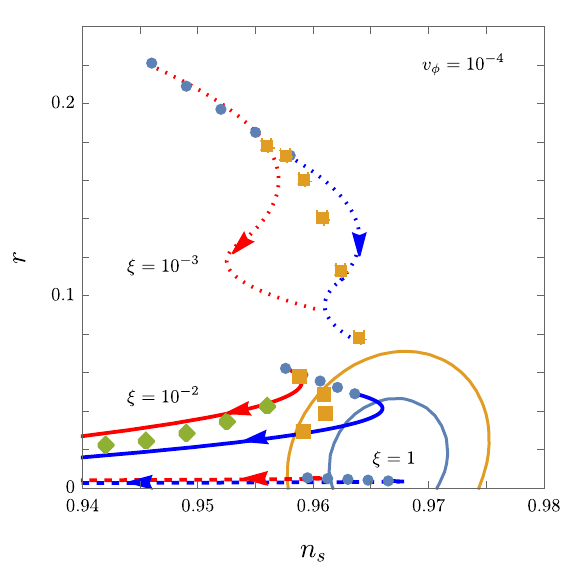}
  \quad
  \includegraphics[width=0.48\textwidth]{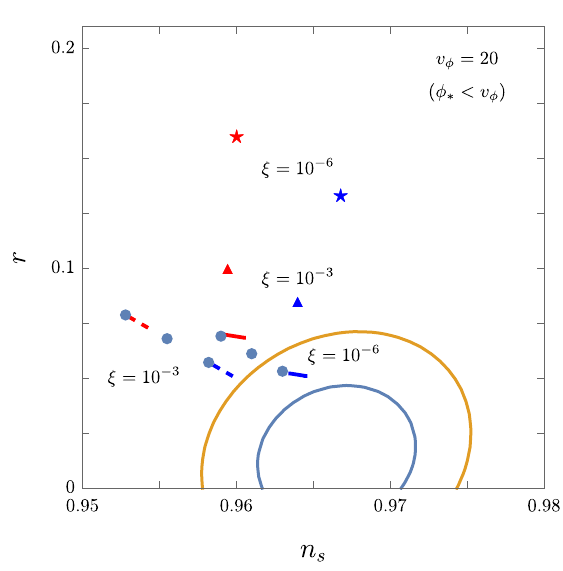}
  \caption{
    Typical predictions for the spectral index and the tensor-to-scalar
    ratio for a fixed value $\chis = \pi / 3$. The red and blue lines
    correspond to $N = 50$ and $N=60$, respectively. The dashed, solid,
    and dotted lines in the left panel correspond to $\xi=1$,
    $10^{-2}$, and $10^{-3}$. The dashed (solid) lines in the right panel
    represent $\xi=10^{-3}$ ($\xi=10^{-6}$). The soft-breaking mass
    ranges are $0\leq m_\chi \leq 1.4\times 10^{-5}$ ($\xi=10^{-3}$),
    $0\leq m_\chi \leq 2.3\times 10^{-5}$ ($\xi=10^{-2}$), and
    $0\leq m_\chi \leq 1.1\times 10^{-4}$ ($\xi=1$) in the left panel,
    and $0\leq m_\chi \leq 7.0 \times 10^{-6}$ in the right panel. The
    arrows indicate the direction of increasing $m_\chi$. The markers
    show fixed values of $m_\chi$: $m_\chi =0$ (circle), $6\times
    10^{-6}$ (square), $7.6 \times 10^{-6}$ (diamond), $5.9 \times
    10^{-6}$ (blue star), $7.0 \times 10^{-6}$ (red star), $6.2\times
    10^{-6}$ (blue triangle), and $7.0 \times 10^{-6}$ (red triangle).
    The contours indicate the $1\sigma$ and $2\sigma$ confidence regions
    from Planck observations.
  }
  \label{Figure:nsr}
  \bigskip
\end{figure}

Figure~\ref{Figure:nsr} shows typical predictions for $(n_s,r)$ in the
present two-field model. The power spectrum is mainly normalized by the
potential parameter $\lambda$ so that it is consistent with the Planck
results. The symmetry-breaking scale $v_\phi$ is chosen as a
representative small or large value: $v_\phi = 10^{-4}$ in the left
panel and $v_\phi = 20$ in the right panel, both for a fixed value
$\chis = \pi /3$. In addition, we take $0\leq m_\chi \leq 1.4\times 10^{-5}$ ($\xi=10^{-3}$), $0\leq m_\chi \leq 2.3\times 10^{-5}$ ($\xi=10^{-2}$), and $0\leq m_\chi \leq 1.1\times 10^{-4}$ ($\xi=1$) in
the left panel, and $0\leq m_\chi \leq 7.0 \times 10^{-6}$ ($\xi=10^{-6}$
and $\xi=10^{-3}$) in the right panel.

We first focus on the left panel of Figure~\ref{Figure:nsr}. When the
soft-breaking mass $m_{\chi}$ is sufficiently small, the Higgs potential
dominates the inflaton potential, and the predictions of the observables
are similar to those of Higgs inflation. In this case, $n_s$ increases
and $r$ decreases as the number of e-folds $N$ increases. This behavior
arises because the field value $\phis$ must be large enough to flatten
the inflaton potential during inflation, leading to larger $n_s$ and
smaller $r$. The variation of $n_s$ becomes more pronounced for larger
$N$, since $m_\chi$ appears in the Lagrangian in combination with
$m_\chi^2 \phi^2$, effectively making $\phis$ larger. As $m_{\chi}$
increases, the inflaton motion in the $\chi$-direction also contributes
to inflation. Consequently, $\phis$ moves closer to the minimum of the
Higgs potential to maintain a fixed number of e-folds. This leads to a
smaller inflaton potential, and thus a smaller $\varepsilon_{*}$ due to
the Planck normalization. By analogy with single-field inflation, this
corresponds to an increase in $n_s$. Therefore, when the pNGB mode also
contributes to inflation with $\phis > v_{\phi}$, $n_s$ tends to
increase. Conversely, for sufficiently large $m_{\chi}$, natural-like
inflation is realized. The left panel of Figure~\ref{Figure:nsr} shows
that larger $m_\chi$ values reduce both $n_s$ and $r$. This feature is
also found in single-field natural inflation. For small $\xi$ and large
$m_{\chi}$, the pNGB part of the action~\eqref{formula:Action} during
inflation can be approximated as
\begin{align}
  S \approx \int d^{4}x \sqrt{-g} \Bigg[ \frac{ \phis^2 }{2}
  \partial_{\mu}\chi \partial^{\mu}\chi -\frac{ m_{\chi}^{2} \phis^{2} }{4} ( 1 - \cos 2\chi ) \Bigg].
\end{align}
This action indicates that when $\phis$ exceeds the Planck scale, the
pNGB potential becomes flatter, favoring natural-like inflation.
Therefore, even for $v_\phi$ smaller than the Planck scale, natural-like
inflation is possible in the present model. A larger $\xi$ further
flattens the inflaton potential, generally leading to a smaller value
of $r$.

Next, we discuss the case with a large symmetry-breaking scale (right
panel of Figure~\ref{Figure:nsr}). When the soft-breaking mass $m_\chi$
is small, increasing $N$ enhances $n_s$ and suppresses $r$. This occurs
because, to maintain the flatness of the potential during inflation, a
larger $N$ requires the inflation to start from a flatter region of the
potential. In this regime, inflation is driven almost entirely by the
radial component $\phi$. When the Higgs potential ceases to dominate,
natural inflation can occur for certain $m_\chi$ values, leading to
larger $r$. This behavior can be seen in the figure, showing that
natural-like inflation also arises for large $v_\phi$, similar to
single-field natural inflation. Finally, regarding the dependence on
$\xi$, when $m_\chi$ is small and the Higgs potential dominates, a
larger $\xi$ tilts the potential in the region $\phi < v_{\phi}$,
resulting in smaller $n_s$ and larger $r$. In contrast, when the pNGB
potential dominates, a larger $\xi$ flattens the inflaton potential,
producing smaller $r$. These trends can be observed in the right panel
of Figure~\ref{Figure:nsr}.

\subsection{Constraints on potential parameters}
\label{Sec:Constraints}

We present the parameter space of the potential couplings $m_\chi$,
$\xi$, and $\lambda$ that are consistent with cosmological
observations. In this paper, we mainly focus on the case
\begin{align}
  \phi > v_\phi \ll 1,
  \label{eq:phi-vphi-condition}
\end{align}
and examine the parameter space and its inflationary features. This
choice realizes a large-field inflation scenario in which the symmetry
breaking becomes relevant at low energies. In the following analysis, we
fix $v_\phi =10^{-4}$ as a representative value for the VEV, a scale
motivated by the grand unification scale and the Majorana mass of right-handed neutrinos. When $v_\phi \ll 1$, inflation with $\phi < v_\phi$ is
not allowed because the slow-roll condition is not satisfied in that
region. Other possible combinations of the symmetry-breaking scale and
the field value $\phi$ are discussed in
Appendix~\ref{sec:inflation-patterns}. The remaining parameter, the
soft-breaking mass $m_\chi$, is analytically examined below for both
the small and large limits.

\paragraph{\underline{Small $m_\chi$ region}:}

In this case, inflation is predominantly driven by the radial component
$\phi$. The spectral index and the tensor-to-scalar ratio are
approximately given by 
\begin{align}
  n_s &\approx 1 - \frac{8 (3 + 5 \xi \phis^2 + 24 \xi^2 \phis^2 + 2 \xi^2 \phis^4 + 12 \xi^3 \phis^4 )}{\phis^2(1 + \xi \phis^2 + 6 \xi^2 \phis^2)^2} + \mc{O} ( m_{\chi}^{2} ),
  \label{Formula:Approns1}
  \\
  r &\approx \frac{128}{\phis^2 ( 1 + \xi \phis^2 + 6 \xi^2 \phis^2)} + \mc{O} ( m_{\chi}^{2} ),
\end{align}
and the e-folding number is expressed as~\cite{Abe:2020ldj}
\begin{align}
  N & \approx \frac{1 - \sqrt{ 1 + 32 \xi + 192 \xi^2} + 2 (1 + 6 \xi ) \xi \phis^2 + 12 \xi \ln \frac{1 + 12 \xi + \sqrt{1 + 32 \xi + 192 \xi^2}}{2 (1 + 6 \xi) ( 1 + \xi \phis^2)}}{16 \xi}.
  \label{Formula:ApproN1}
\end{align}
When $\xi$ is also small, these expressions reduce to those of chaotic
inflation:
\begin{align}
  n_s \approx 1 - \frac{3}{N} + \mc{O}(m_\chi^2),
  \qquad 
  r \approx \frac{16}{N} + \mc{O}(m_\chi^2).
\end{align}
This parameter region is excluded by observations because a small $\xi$
makes the inflaton potential steeper, driving spacetime farther from the
de Sitter limit and leading to $r$ values larger than observed. From the
above discussion and approximate expressions, we conclude that there is
a lower bound on the non-minimal coupling $\xi$~\cite{Abe:2020ldj}. 

The running of the spectral index is approximately given by 
\begin{align}
  \alpha \approx - \frac{3}{N^2} + \mc{O} ( m_{\chi}^{2} ),
  \label{Formula:ApproalphaA}
\end{align}
which is consistent with observations for sufficiently large $N$. This
hierarchy between $|1-n_{s}|$ and $|\alpha|$ is typical in single-field
inflation~\cite{vandeBruck:2016rfv}. For large $\xi$, we obtain the
approximate form
\begin{align}
  \alpha& \approx - \frac{ 32 }{ 9 \xi^2 \phis^4 }.
  \label{Formula:Approalpha1}
\end{align}
A large $\xi$ ensures that $\alpha$ remains small and consistent with
data, again showing a clear hierarchy between $|1 - n_{s}|$ and
$|\alpha|$. Inflation is then effectively dominated by the Higgs
potential and approximated by a single-field scenario.

We next consider the isocurvature fraction for small $m_\chi$. The
general form is given by \eqref{formula:betaiso}, with the transfer
function $T_\mathcal{SS}$ expressed as
\begin{align}
  T_\mathcal{SS} = \exp \int_{N_*}^N \tilde{\beta} \, dN'.
  \label{Formula:betaisoAppro}
\end{align}
The explicit form of $\tilde{\beta}$ is provided in
Appendix~\ref{app:observables}, where it is determined by the model
parameters and field values obtained from the EOMs. In the present case,
$\tilde{\beta}$ can be approximated as
\begin{align}
  \tilde{\beta} \approx& -\frac{8}{\phi^2} \quad ( \xi \ll 1 ), 
  \label{Formula:beta1} \\
  \tilde{\beta} \approx& -\frac{ 4 v_{\phi}^2 }{ 3 \phi^2 } \quad (\xi \gg 1).
  \label{Formula:beta11}
\end{align}
A negative, moderate $\tilde{\beta}$ leads to an exponentially small
$T_\mathcal{SS}$, resulting in a sufficiently small isocurvature
fraction $\beta_{\mr{iso}}$ consistent with observations. Note that
$\xi \gg 1$ corresponds to $\phis \ll 1$.

\paragraph{\underline{Large $m_\chi$ region}:}

\begin{figure}
  \centering
  \begin{minipage}{0.32\textwidth}
    \centering
    \includegraphics[width=\textwidth]{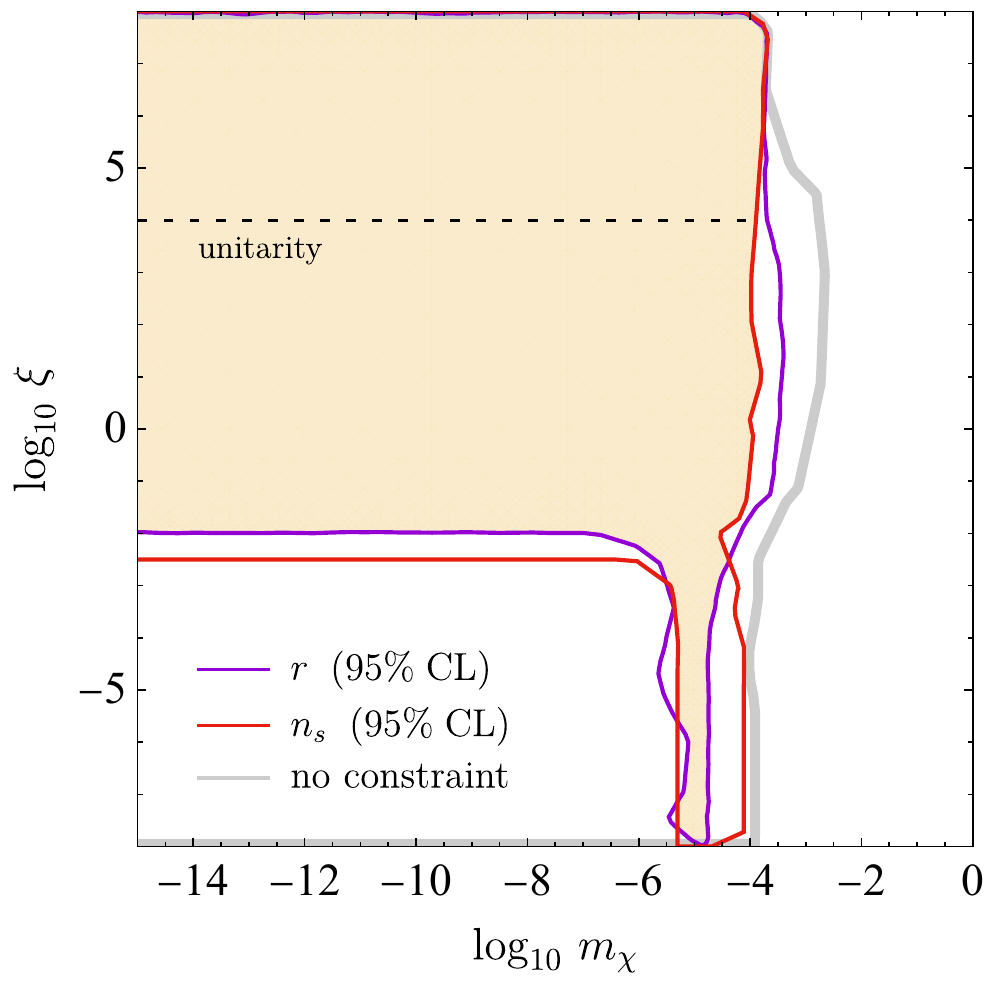}
    \\
    \vspace{2ex}
    \includegraphics[width=\textwidth]{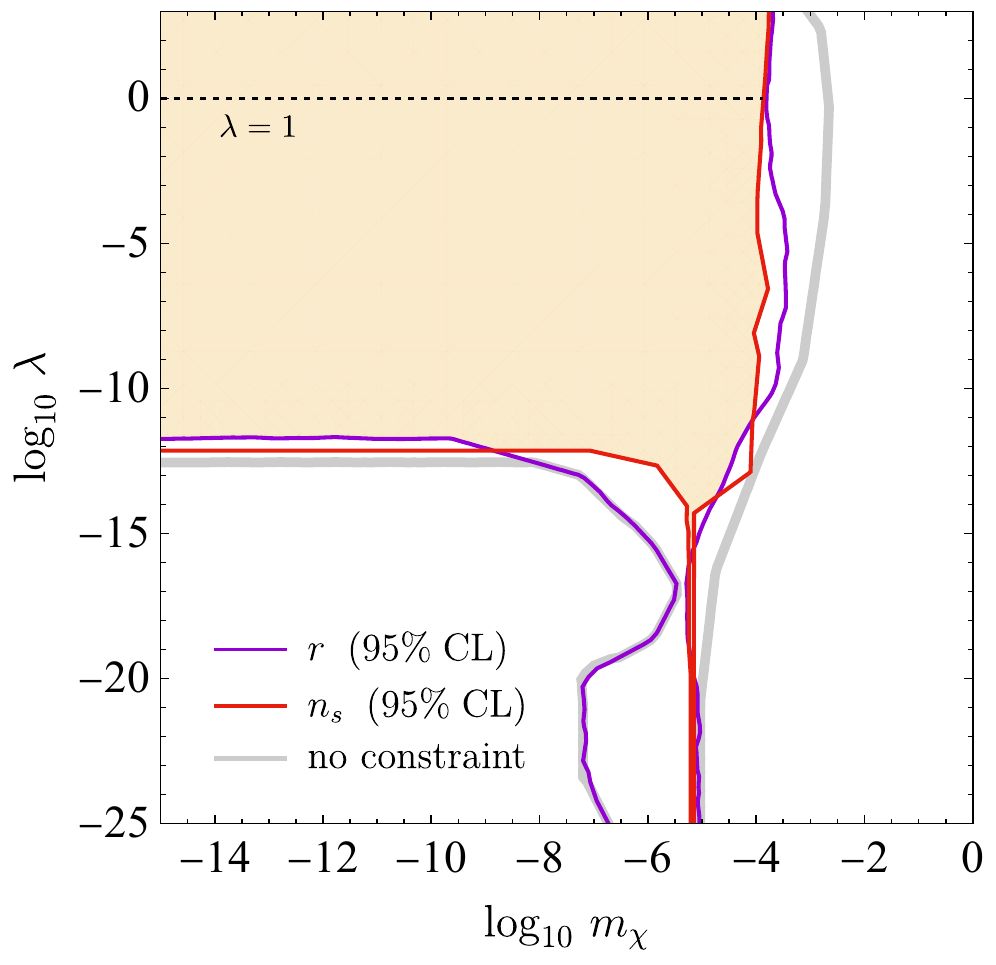}
    \\
    \vspace{2ex}
    \includegraphics[width=\textwidth]{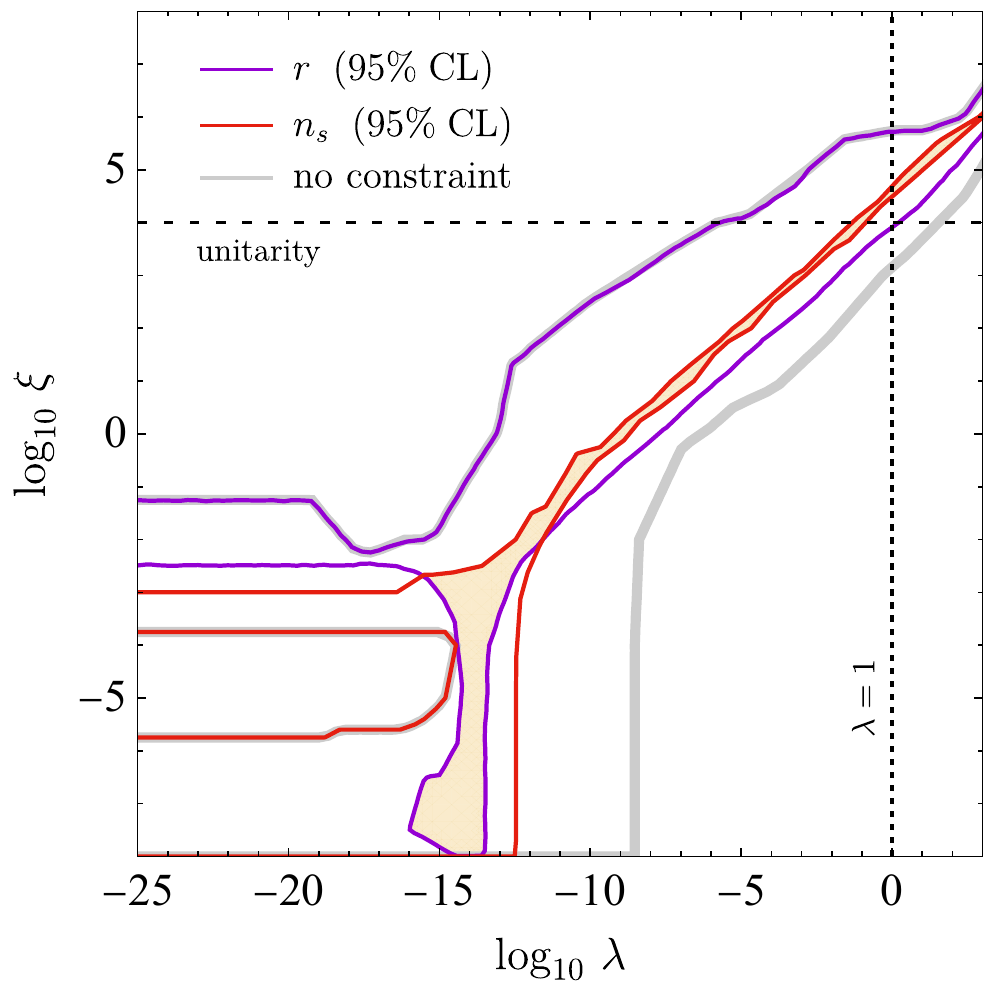}
    \subcaption{Higgs}
  \end{minipage}
  \
  \begin{minipage}{0.32\textwidth}
    \centering
    \includegraphics[width=\textwidth]{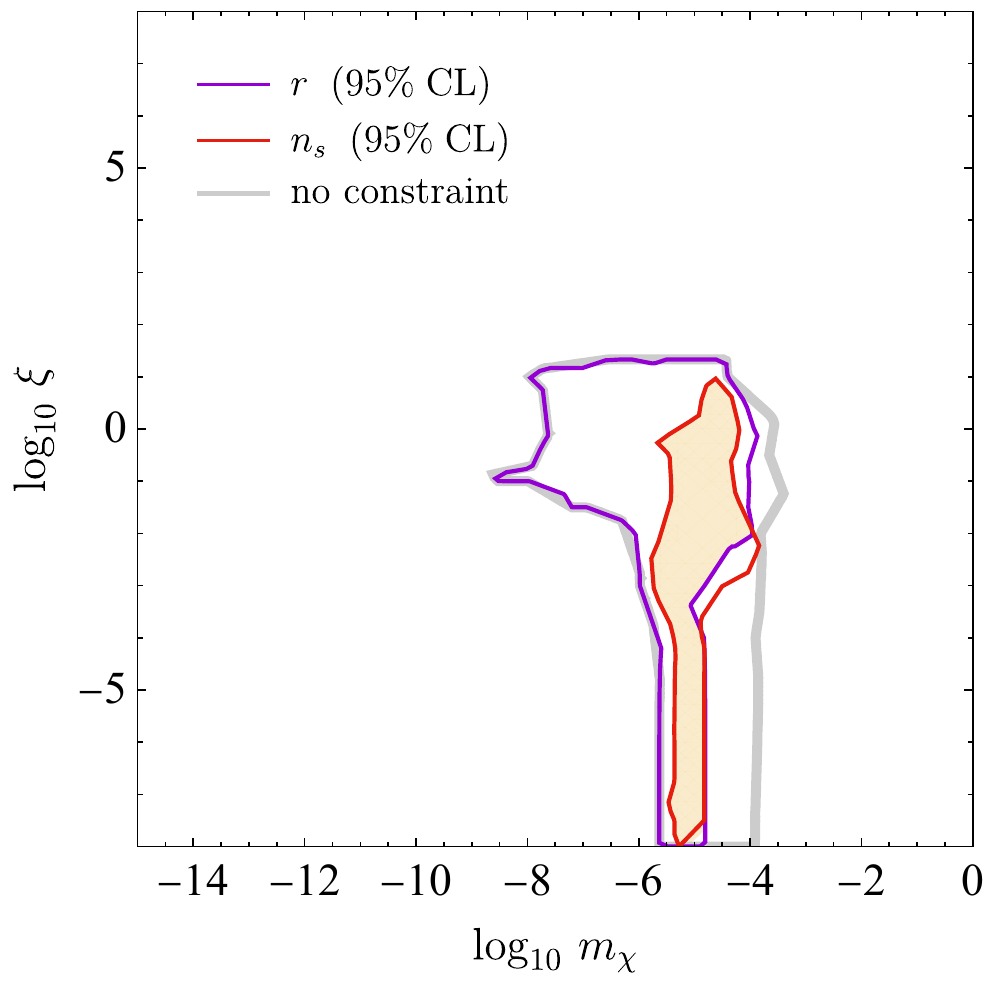}
    \\
    \vspace{2ex}
    \includegraphics[width=\textwidth]{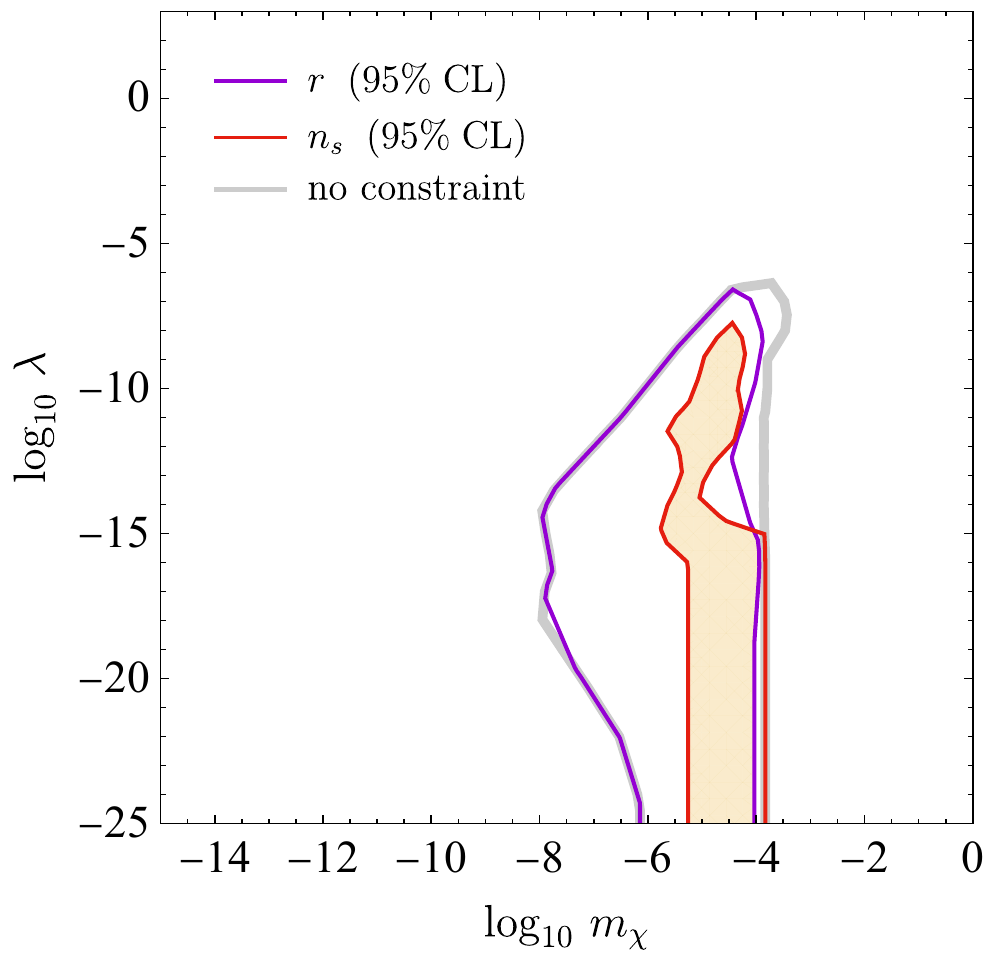}
    \\
    \vspace{2ex}
    \includegraphics[width=\textwidth]{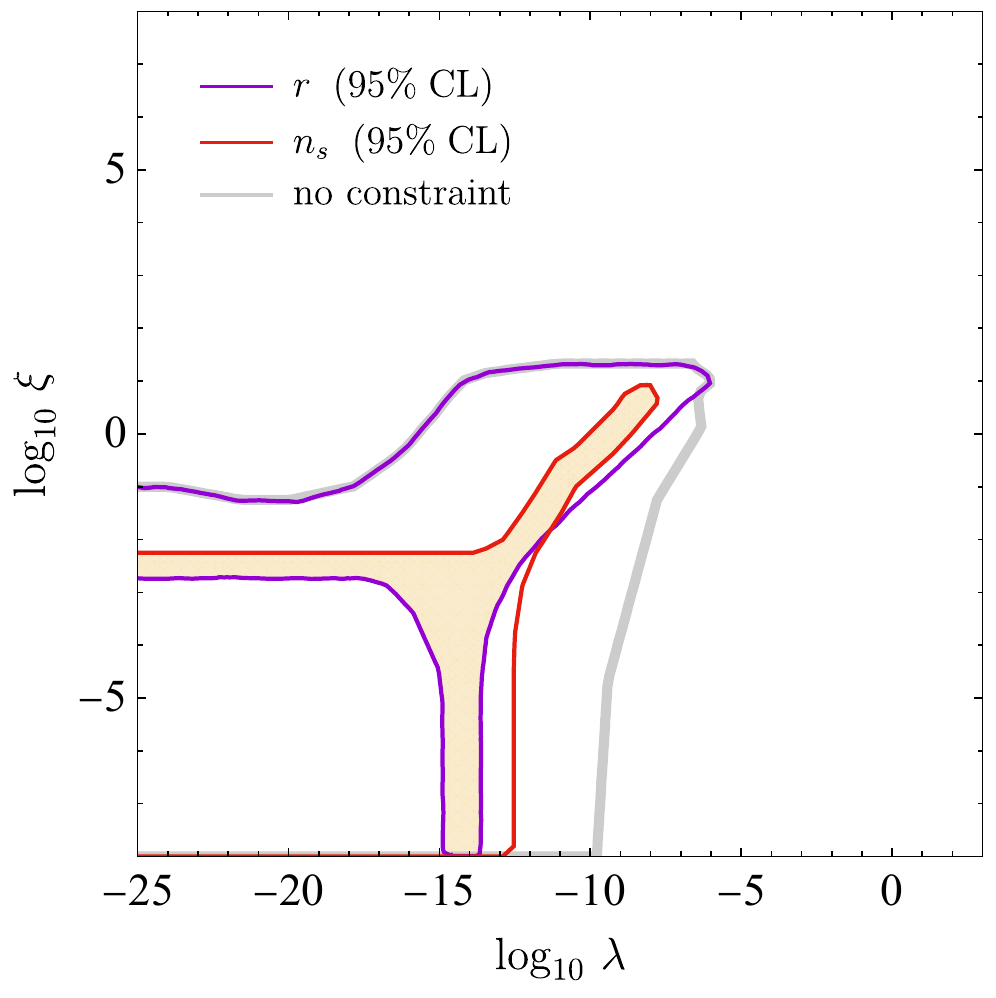}
    \subcaption{Mixed}
  \end{minipage}
  \ 
  \begin{minipage}{0.32\textwidth}
    \centering
    \includegraphics[width=\textwidth]{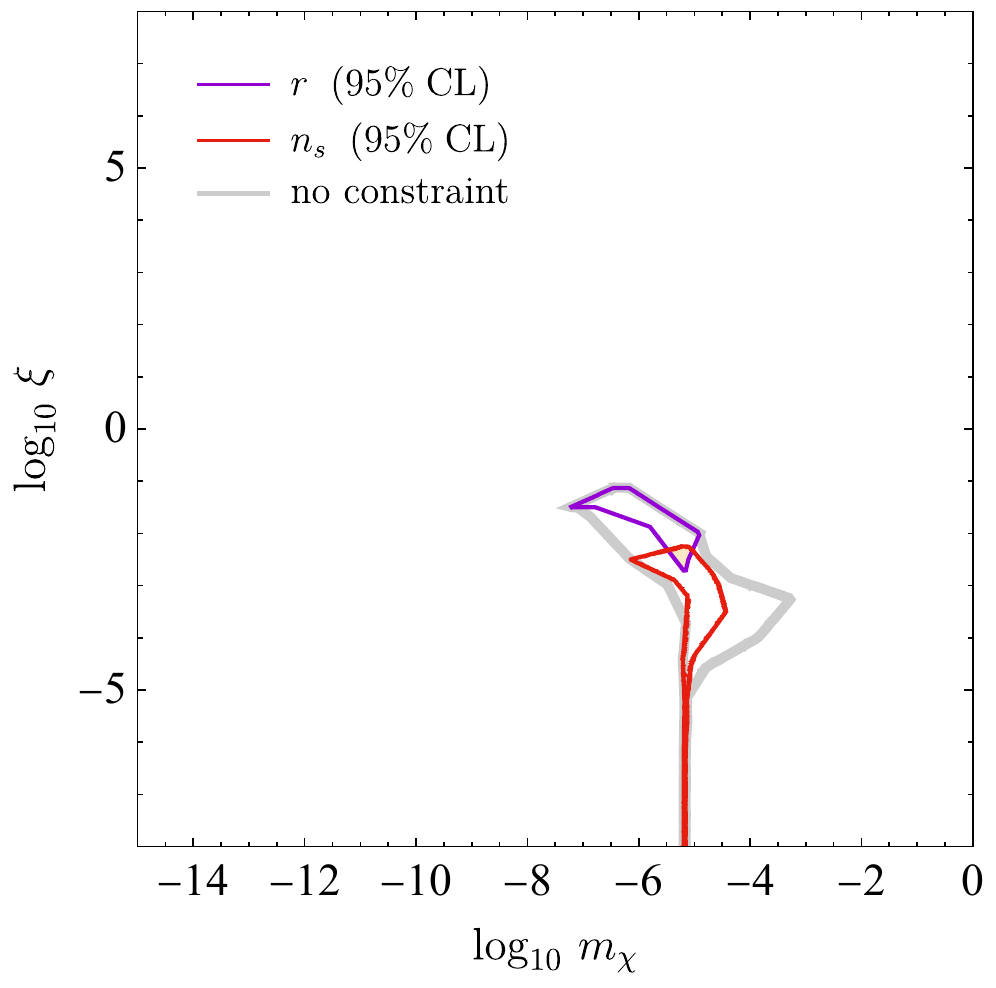}
    \\
    \vspace{2ex}
    \includegraphics[width=\textwidth]{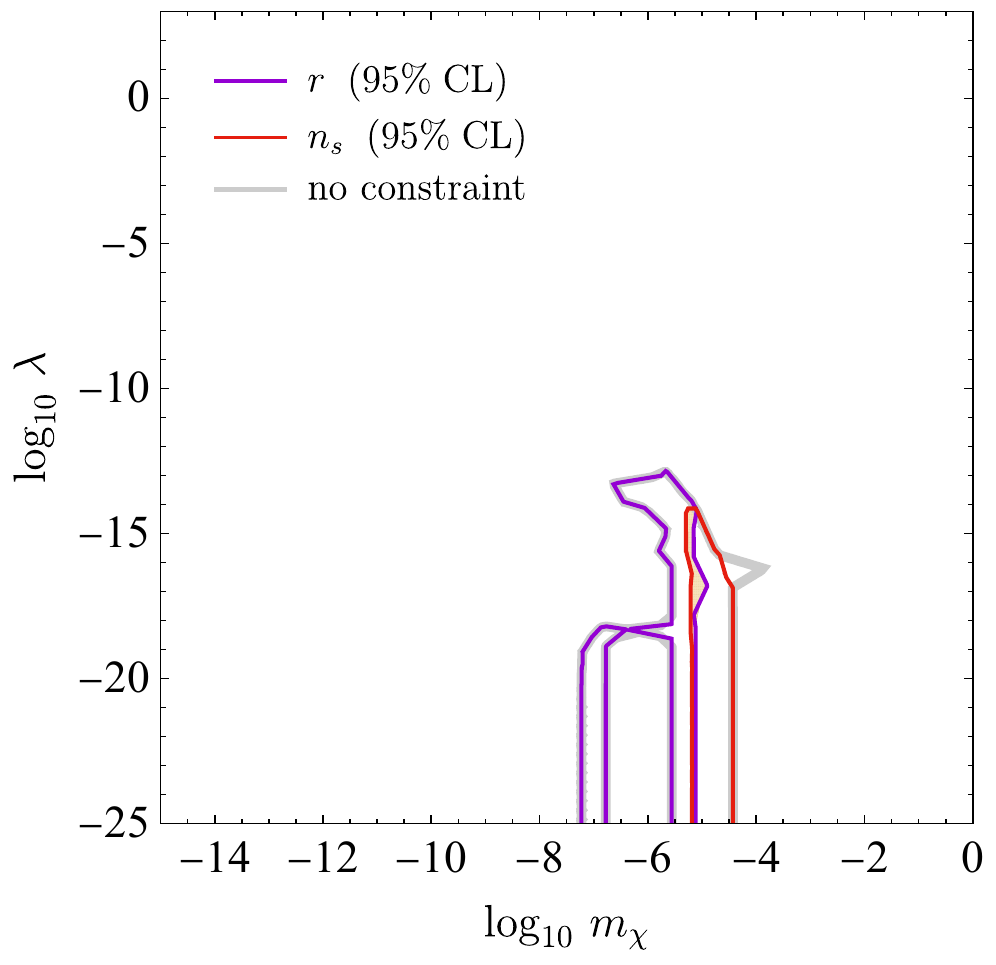}
    \\
    \vspace{2ex}
    \includegraphics[width=\textwidth]{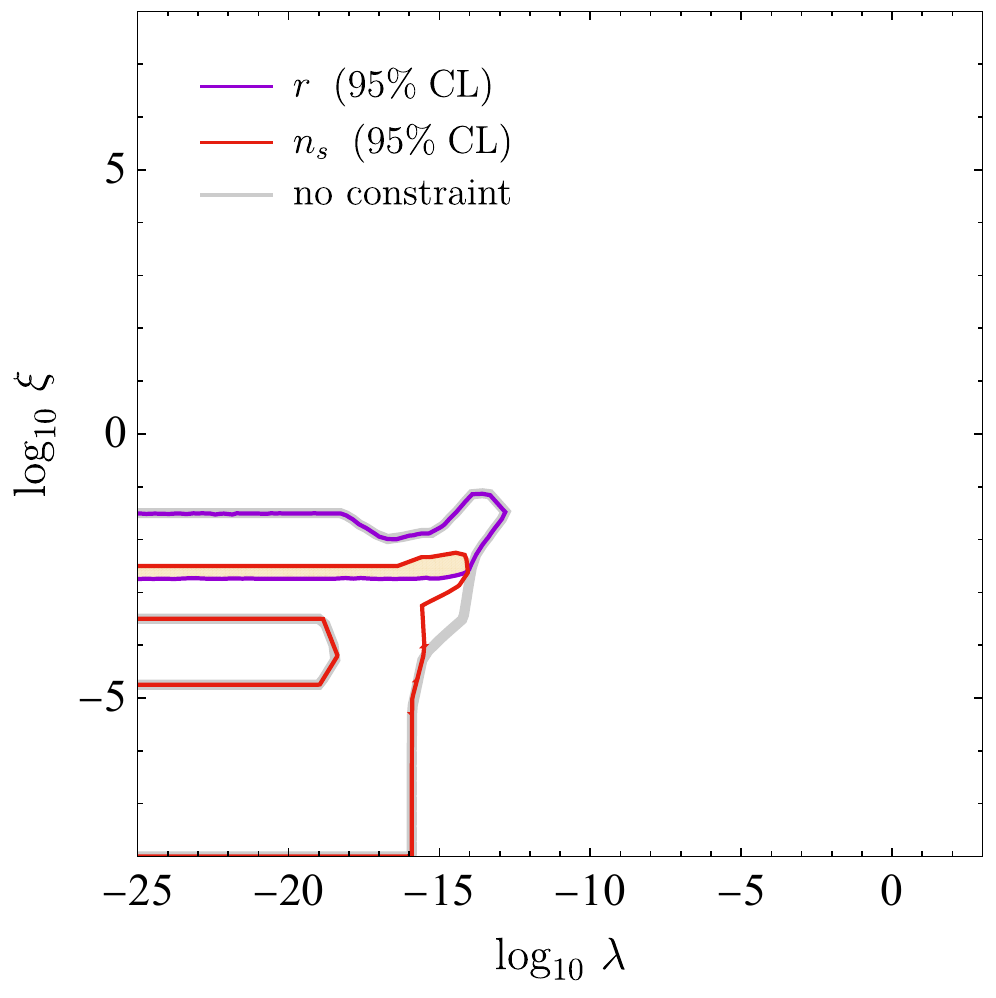}
    \subcaption{Natural}
  \end{minipage}
  \caption{Allowed parameter spaces of the present two-field model.
    The plots in the left, middle, and right columns correspond to
    Higgs, Mixed, and Natural inflation types. The light orange
    regions are consistent with cosmological observations. The
    purple and red lines represent subspaces satisfying the Planck
    constraints only on $r$ and $n_s$, respectively. The regions
    satisfying only the Planck normalization are outlined in gray. The bounds for unitarity and perturbativity on the model parameters are shown in the Higgs-type panels as dashed and dotted lines, respectively.}
  \label{fig:HMNplots}
  \bigskip
\end{figure}

In this case, the $\chi$ component contributes significantly, and both
$\phi$ and $\chi$ play roles as inflatons. For the spectral index, $n_s$
increases with $m_\chi$ when $\xi \ll 1$, as shown in
Figure~\ref{Figure:nsr}, leading to disagreement with observations. In
contrast, regions where inflation is primarily driven by the $\chi$
component can be allowed. In particular, the spectral index can be
approximated as 
\begin{align}
  n_s \approx 1 - \frac{4(2 + T_{\mc{RS}}^{2})}{(1 + T_{\mc{RS}}^{2} )\phis^{2}\sin^2\chis },
\end{align}
for $m_\chi \gg \sqrt{\lambda} v_\phi$ and $\xi \ll 1$. The spectral
index along this trajectory matches the observations. This type of inflation
is driven by the pNGB mode assisted by the dynamical radial component.
Inflation can also occur for $\xi \gg 1$, where the pNGB mode
contributes sufficiently, as shown in the left and middle panels of
Figure~\ref{fig:HMNplots}. This is because a large $\xi$ not only
flattens the potential but also enhances motion in the $\chi$ direction.
However, when the initial value $\chis$ is sufficiently small, the
inflaton rarely moves in the $\chi$ direction, regardless of the
symmetry-breaking scale.

The isocurvature bound can also be discussed. For large $m_\chi$, the
function $\Tilde{\beta}$ in Eq.~\eqref{Formula:betaisoAppro} is approximated as 
\begin{align}
  \Tilde{\beta} &\approx - \frac{ 2 }{ \phi^2 \sin^{2}\chi} \quad (\xi\ll1), 
  \\
  \Tilde{\beta} &\approx - \frac{2\xi}{ \sin^2 \chi} \quad (\xi\gg1),
\end{align}
for $m_\chi \gg \sqrt{\lambda}v_\phi$ and $v_\phi \ll 1$. Since
$\Tilde{\beta}$ is negative in both cases, $\beta_{\mathrm{iso}}$ becomes
sufficiently small for large $m_\chi$.

As discussed in Section~\ref{sec:classification}, the inflaton
trajectories are classified into three types: Higgs, Mixed, and Natural.
In all parameter-constraint plots, we impose the Planck normalization
$\mc{P}_\mc{R} = 2.1 \times 10^{-9}$. 
Figure~\ref{fig:HMNplots} shows the parameter spaces in the $(m_\chi,\xi)$, $(m_\chi, \lambda)$, and $(\lambda,\xi)$ planes for these three
types. In our two-field model under
the condition \eqref{eq:phi-vphi-condition}, experimentally allowed regions
exist for all three inflationary scenarios, shown in light colors. We
impose the constraints from \eqref{eq:obs-constraints}, with purple lines
for $r$ and red lines for $n_s$. 
In the left panels, inflation is
successful over a wide region because a large non-minimal coupling
($\xi \gtrsim 10^{-2}$) consistently flattens the scalar potential along
the $\phi$ direction. The Planck normalization and the perturbativity of
the quartic coupling $\lambda$ impose an upper bound
$\xi\lesssim10^5$~\cite{Bezrukov:2007ep,Rubio:2018ogq,Pareek:2023die}.
In addition, inflation with a non-minimal coupling is known  
to exhibit a unitarity cutoff at $M_P/\xi$, as inferred from  
graviton–scalar-field scattering analyses~\cite{Burgess:2009ea,Barbon:2009ya,Hertzberg:2010dc}. 
Accordingly, we require $H_{\mathrm{inf}} < M_P/\xi$ to ensure  
that inflation proceeds within the regime of validity of the  
effective field theory. This condition leads to the upper 
bound $\xi < 10^4$, which corresponds to the dashed lines in
Figure~\ref{fig:HMNplots}.
For small $\xi$ ($\xi \lesssim 10^{-2}$), the model tends toward chaotic
inflation, which is experimentally excluded. However, an allowed region
appears due to the additional pNGB contribution to inflation, as shown
in the mixed-type middle panels, around $m_\chi \sim 10^{-5}$. The
soft-breaking mass scale is estimated from the Planck normalization as 
\begin{align}
    \mc{P}_{\mc{R}} \approx \frac{V^3}{12 \pi^2 \del_\chi V} \sim
    \frac{m_\chi^2 \phi^2}{12 \pi^2} \approx 2.1 \times 10^{-9}.
\end{align}
A small allowed region for the Natural type also appears near
$m_\chi\sim10^{-5}$ and $\xi\sim 10^{-3}$, where inflation is mainly
driven by $\chi$, and $\phi$ oscillates at the end of inflation. The
inflationary dynamics in this region resembles those discussed in
Refs.~\cite{McDonough:2020gmn,Lorenzoni:2024krn}. The natural-type
inflation can thus be regarded as a limiting case of the mixed type in
this analysis.

We thus find a variety of parameter regions consistent with
cosmological observations. All three inflationary types, classified by
their trajectories, possess experimentally allowed parameter regions. In
the next section, we investigate the reheating and the generation of
the matter-antimatter asymmetry as further cosmic phenomena in the
viable parameter space.

\medskip

\section{Cosmic reheating}
\label{sec:reheating}

In this section, we introduce RH neutrinos that couple via Yukawa
interactions to the complex scalar field $\Phi$, and analyze
the reheating~\cite{Abbott:1982hn,Kolb:1990vq}, whereby the inflaton
transfers its energy to the thermal bath through the decay into the
neutrino sector. In the vacuum after the inflation epoch, the VEV $v_\phi$
generates RH-neutrino masses and a pNGB associated with the lepton-number
$\rU(1)_L$ symmetry, known as the majoron~\cite{Chikashige:1980ui,Schechter:1981cv,Gelmini:1980re,Rothstein:1992rh}. The majoron has been
widely studied in dark-matter physics~\cite{Berezinsky:1993fm,Matsumoto:2010hz,Frigerio:2011in,Queiroz:2014yna,Garcia-Cely:2017oco,Brune:2018sab,Abe:2020dut,Manna:2022gwn,Biggio:2023gtm,Akita:2023qiz,Bonilla:2023egs} and in scenarios for generating lepton
asymmetry~\cite{Pilaftsis:2008qt,LeDall:2014too,Alanne:2018brf,Abe:2021mfy,Barreiros:2022fpi}. Here we show that these RH-neutrino
couplings can also mediate the reheating for a non-minimally coupled inflaton.

\subsection{Coupling to RH neutrinos}

We consider the following action for the RH neutrinos $N_i$ in the
Einstein frame,
\begin{align}
  S_{N} = \int d^4x \sqrt{-g} \biggl[ \frac{1}{2 ( 1 + 2 \xi |\Phi|^2 )^\frac{3}{2}} \ol{N_i} i \sld{D} N_i - \frac{1}{2 ( 1 + 2 \xi |\Phi|^2 )^2} (f_i\Phi \ol{N_i}  P_R N_i +\text{h.c.})\biggr],
  \label{Formula:ActionRH1}
\end{align}
where $f_i$ are generation-diagonal Yukawa couplings, in general complex.
The covariant derivative $\slashed{D}$ includes the spin connection, and
$P_R$ is the chiral projection operator. This form, in particular the
$\Phi$ dependence of the coefficients, follows from the canonical
Jordan-frame action for $N_i$ after the Weyl rescaling by the inflaton
field, analogously to Section~\ref{subsec:Model}. The action respects
the lepton number, under which $N_i$ and $\Phi$ carry charges $+1$ and $-2$,
respectively. This symmetry corresponds to the U(1) phase rotation of
the inflaton field (the origin of the pNGB and its soft-breaking mass),
naturally motivating a two-field inflation framework with a complex
scalar. In what follows, we focus on the case where the $\phi$ field
oscillates at the end of inflation and decays into the thermal bath via
RH neutrinos. We assume $\chi$ stays at a constant value $\bar{\chi}$
during reheating and perform the chiral rotation
\begin{align}
  N_i \to e^{-i \gamma^5 (\bar{\chi}+\arg f_i) /2} N_i,
\end{align}
so that the Yukawa couplings (and hence the RH-neutrino masses) are
real. During the reheating era, the inflaton oscillates about the
minimum of the potential determined by the VEV $u_\phi$,
\begin{align}
  u_\phi = \frac{\Re\big(\lambda v_\phi^2\Omega^2(v_\phi)-2m_\chi^2\sin^2\bar\chi\big)^{1/2}}{\big(\lambda\Omega^2(v_\phi)-2\xi m_\chi^2\sin^2 \bar{\chi}\big)^{1/2}}.
  \label{Formula:uphi}
\end{align}
We here do not consider parameter regions where the vacuum has the
runaway behavior ($u_\phi\to\infty$). Note that $u_\phi$ can vanish if
the pNGB potential dominates during the reheating. Expanding the
scalar field around this VEV, we canonically normalize the kinetic term as
\begin{align}
  \phi = u_\phi+\frac{\Omega^2(u_\phi)}{\sqrt{\Omega^2(u_\phi)+6\xi^2 u_\phi^2}}\rho,
\end{align}
and obtain the action for the scalar fluctuation and the canonically
normalized RH neutrinos:
\begin{align}
  S &= \int d^4x \sqrt{-g} \Biggl[
      \frac{1}{2} \del_\mu \rho \del^\mu \rho -V(\rho)
      + \frac{1}{2} \ol{N_i} (i \sld{D} -m_{N_i})N_i 
      - \frac{1}{2} y_i \rho \ol{N_i} N_i
      \Biggr].
  \label{Formula:ReheatingAction}
\end{align}
The inflaton–RH-neutrino coupling $y_i$ and the mass eigenvalues
$m_{N_i}$ and $m_\rho$ are
\begin{align}
  y_i &= \frac{\Omega(u_\phi)}{\sqrt{\Omega^2(u_\phi)+6\xi^2 u_\phi^2}}\frac{f_i}{\sqrt{2}},   \\
  m_{N_i} &= \frac{1}{\Omega(u_\phi)}\frac{f_i}{\sqrt{2}} u_\phi,
  \label{Formula:massofN} 
  \\
  m_\rho &=  \frac{\sqrt{\lambda \Omega^2(v_\phi)\big[u_\phi^2(3+5\xi v_\phi^2-3\xi u_\phi^2)-v_\phi^2\big]   +2m_\chi^2(1-8\xi u_\phi^2+3\xi^2u_\phi^4)\sin^2\bar\chi}}{\sqrt{2}\,\Omega^2(u_\phi)\sqrt{\Omega^2(u_\phi)+6\xi^2 u_\phi^2}}.
  \label{Formula:massofrho}
\end{align}

\subsection{Reheating and leptogenesis}

We now examine whether the inflaton coupling to RH neutrinos can be
consistent with the reheating of the universe after inflation. A
simplified analysis is performed by taking into account rough
estimations of the seesaw-induced neutrino masses and the conditions
required for successful leptogenesis.

The inflaton decays into RH neutrino pairs, $\rho \to N_i N_i$, during
its oscillation. The decay width is determined by the Yukawa coupling $y_i$:
\begin{align}
  \Gamma_{\rho \to N_iN_i} = \frac{y_i^2}{16\pi} m_\rho \bigg(
  1 - \frac{4 m_{N_i}^2}{m_\rho^2}\bigg)^{3/2}.
  \label{Formula:GammarohNN}
\end{align}
Once the energy is transferred to the $N_i$ sector, the Standard Model
(SM) fields become thermalized through the decay and scattering of $N_i$
if the neutrino Yukawa couplings $y_{\nu_i}$ are sufficiently large. The
decay process $N_i \to L_jH$ (with $L_j$ and $H$ denoting the lepton and
Higgs fields, respectively) is effective when $N_i$ is heavier than the
electroweak scale and its decay rate exceeds the Hubble parameter at
temperature $T$. This condition can be written in terms of the model
parameters as
\begin{align}
  T < 21\, g_*^{-1/4}y_i v_\phi
  \Big(\frac{m_{\nu_i}}{10^{-1}\,\text{eV}}\Big)^{1/2}
  \eqqcolon T_D ,
  \label{eq:TD}
\end{align}
where $g_*$ denotes the effective number of relativistic degrees of
freedom in the radiation bath that defines the temperature. The
seesaw-induced light neutrino mass is given by
$m_{\nu_i} = (y_{\nu_i} v_h)^2 / m_{N_i}$ with $v_h = 175$~GeV (the
generation structure is omitted for simplicity). The scattering of $N_i$
to the SM sector is also effective if its rate is sufficiently large.
Such a process can contribute to the thermalization of the SM radiation
component when the decay process is inefficient. A rough estimation of
its efficiency can be obtained by evaluating the collision term from the
scattering, which leads to
\begin{align}
  T < 0.17\, g_*^{-3/2} y_i v_\phi
  \Big(\frac{m_{\nu_i}}{10^{-1}\,\text{eV}}\Big)
  \eqqcolon T_S .
  \label{eq:TS}
\end{align}
We have approximated the effective degrees of freedom for the entropy
density as equal to those for the energy density. The SM sector decouples
from the scattering process when its temperature exceeds $T_S$.

\begin{figure}[t]
  \centering
  \includegraphics[width=0.31\linewidth]{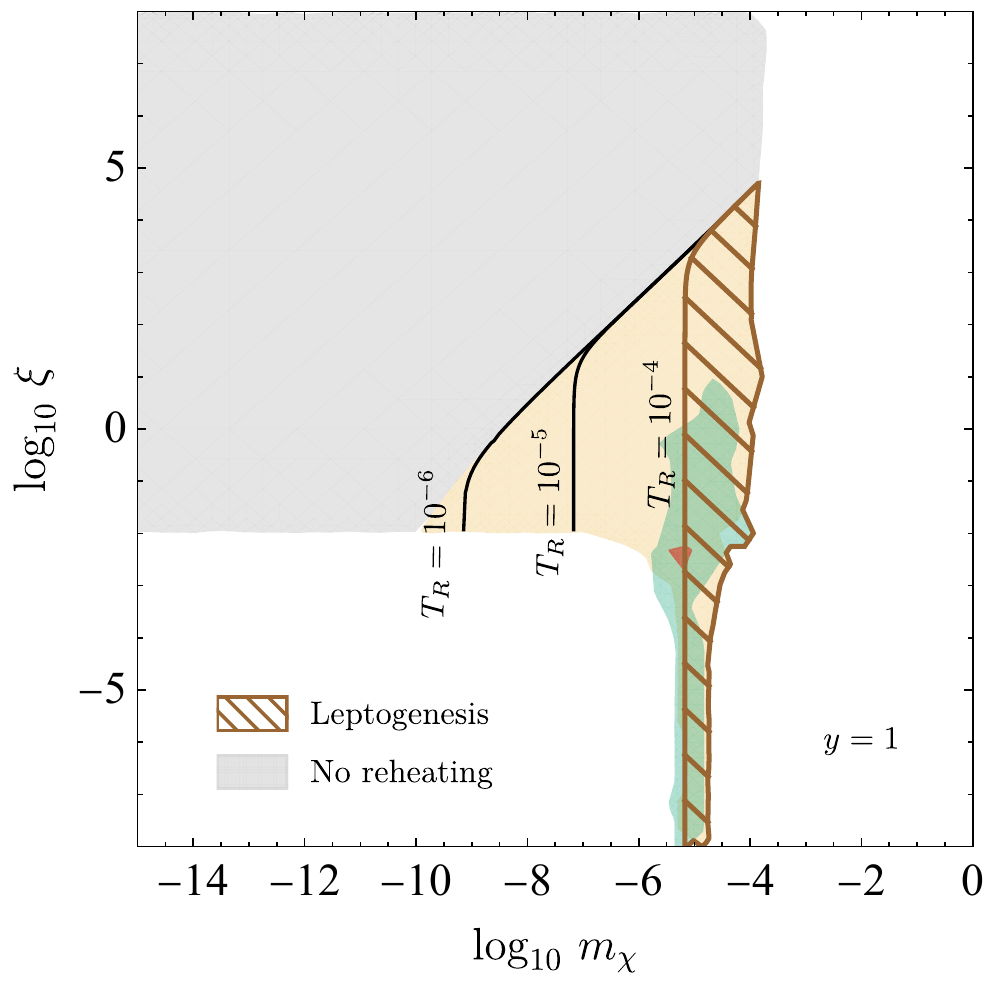}
  \quad
  \includegraphics[width=0.31\linewidth]{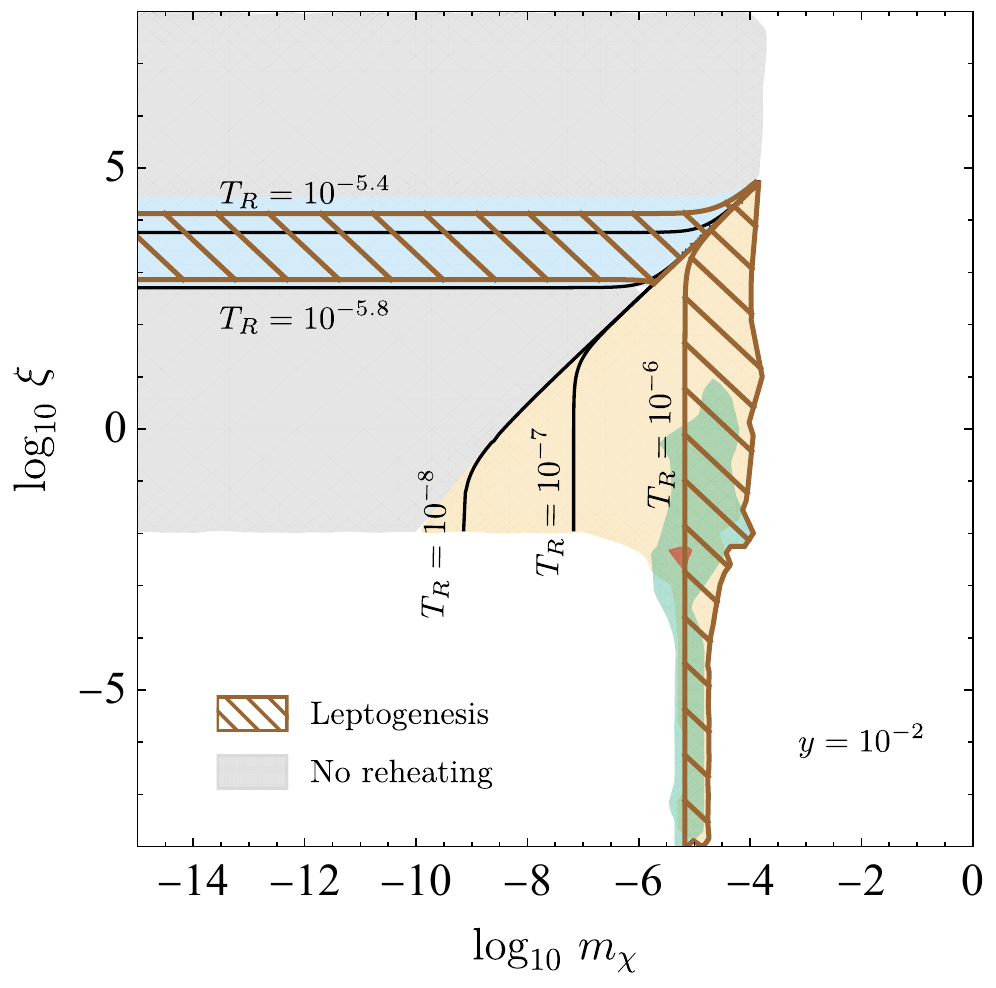}
  \quad
  \includegraphics[width=0.31\linewidth]{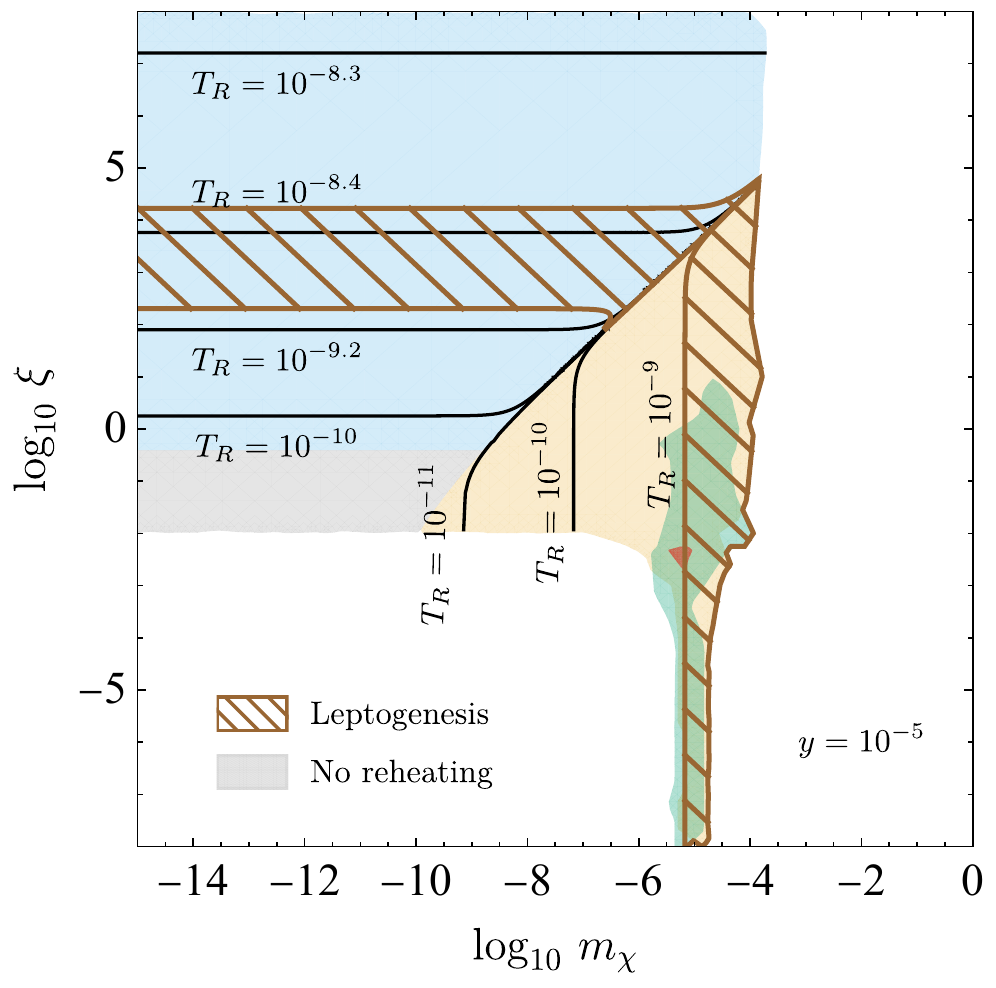}
  \caption{
    Parameter space for inflation and lepton number generation in the
    present two-field model. Typical reheating temperature $T_R$ from
    the inflaton decay into RH neutrinos with couplings $y=1$, $10^{-2}$,
    and $10^{-5}$. We show $T_R$ on the parameter space consistent with
    the inflationary observables. The lepton number can be successfully
    generated in the brown hatched region, and the reheating does not occur
    in the gray region. For the meaning of other colors, see the text.}
  \label{fig:LGyukawa}
  \bigskip
\end{figure}

As a representative example of this type of reheating, we focus on the
parameter region consistent with the inflationary observables shown in
the upper panels of Figure~\ref{fig:HMNplots}.
Figure~\ref{fig:LGyukawa} presents the classification of reheating and
leptogenesis in the inflationary-allowed region of the
$(m_\chi,\xi)$-plane for the Yukawa couplings $y=1$, $10^{-2}$, and
$10^{-5}$. In the gray and blue regions, the scalars oscillate around
the symmetry-breaking vacuum, and the inflaton decay channel into RH
neutrinos is open only in the blue region. The yellow, green, and red
regions realize consistent inflation and reheating at the trivial
vacuum. These colors correspond to the three types of inflaton
trajectories—Higgs, Mixed, and Natural, respectively. As shown below,
successful lepton number generation occurs in the hatched brown region.

Here we comment on the magnitude of the Yukawa coupling $y$, in
particular its upper bound. Introducing the coupling $y$ may induce
corrections to the scalar-field EOMs used in the previous section. In
the following analysis, we assume that these corrections are small and
neglect them. Specifically, for the friction term, a rough estimate of
the consistency condition is $\Gamma < H$, where $\Gamma$ is evaluated
using the effective mass $V''$, leading to the inequality condition 
$V''(y^2/8\pi)^2 < V$. In two-field inflation models, various
trajectories exist, making the effective mass evaluation complicated.
While the mass scale is usually restricted by the slow-roll argument in the
single-field case, for scalar directions not responsible for inflation a
large mass can coexist with the coupling $y$. Setting this large mass to
the Planck scale gives the most stringent constraint, which translates
into $y < \sqrt{8\pi} V^{1/4} \sim \mathcal{O}(10^{-1})$. In the
numerical analysis below, we take representative values $y=1$, $10^{-2}$
and $10^{-5}$, noting that for $y=1$ certain constraints on the
potential parameters arise. In practice, the constraint appears
satisfied if $m_\chi < 8\pi \sqrt{V} \sim \mathcal{O}(10^{-3})$.

When the inflaton settles into the trivial vacuum $u_\phi=0$ after the
inflationary expansion, it oscillates around the origin, and the decay
process becomes inefficient if the RH neutrino mass is regarded as 
small. According to \eqref{eq:TS}, the SM sector is thermalized up to
the temperature $T_S$ with $g_*=112$, at which it decouples from the
thermal bath. Since the decay width \eqref{Formula:GammarohNN} is large
enough, the RH neutrino sector is subsequently heated to the temperature
\begin{align}
  T_N = 0.17\, g_*^{-1/4} y_i m_\rho^{1/2},
  \label{eq:TN}
\end{align}
with $g_*=5.25$. Here $T_N$ is defined by the condition that the Hubble
time equals the decay width \eqref{Formula:GammarohNN}. (If
$T_N<T_S$, the $N_i$ reheating phase does not occur, and the reheating
temperature is given by $T_N$.) As soon as this reheating is over, $\rho$ is
assumed to roll immediately to the nontrivial vacuum, where the RH
neutrinos become massive and their decays heat the SM sector. If
$T_N<T_D$, the SM sector is promptly thermalized; otherwise,
the SM thermalization takes place when the universe cools down to $T_D$. The
final reheating temperature $T_R$ of the SM sector is then expressed as
\begin{align}
  T_R = \begin{cases}
     T_N & (T_N < T_S), \\
     \bigl((1-\omega)T_S^3+\omega T_N^3\bigr)^{1/3} & (T_S < T_N < T_D),\\
     \bigl((1-\omega)T_S^3+\omega T_D^3\bigr)^{1/3} & (T_S < T_D < T_N),
   \end{cases}
\end{align}
where $\omega \coloneqq g_*^{3N}/(g_*^{\mathrm{SM}}+g_*^{3N})$. In the
region of interest shown in Figure~\ref{fig:LGyukawa}, the relation
$T_S<T_N<T_D$ is numerically satisfied everywhere, yielding 
the reheating temperature of the radiation 
$T_R\simeq0.36T_N\simeq0.027\,y_i m_\chi^{1/2}$. The parameter
dependence of $T_R$ on $y_i$ and $m_\chi$ is visible in the purple
regions of Figure~\ref{fig:LGyukawa}.

We now comment on the leptogenesis, i.e., lepton number generation through
the out-of-equilibrium decay of RH neutrino~\cite{Fukugita:1986hr}. The
mechanism can work if at least $T_R>m_{N_i}$ is satisfied. This
condition translates into 
$0.027\,m_\chi^{1/2}>v_\phi$, implying $m_\chi>1.3\times10^{-5}$ for the
parameters in Figure~\ref{fig:LGyukawa}. The condition is 
independent of the Yukawa couplings and common to all three panels.
That indicates a small portion of the inflationary-allowed region is
compatible with successful reheating and leptogenesis.

The other case corresponds to a non-vanishing VEV $u_\phi$, where the
inflaton oscillates around the nontrivial vacuum. In
Figure~\ref{fig:LGyukawa}, this corresponds to the blue and gray
regions, and numerically $u_\phi$ is found to be close to $v_\phi$. 
In this case, the RH neutrinos are massive, and the reheating via the
inflaton decay into $N_i$ is possible if $m_\rho>2m_{N_i}$. This minimum
requirement for reheating imposes the following parameter constraint
\begin{align}
  \Omega^2(v_\phi)\lambda^{1/2}
  > 2\bigl(\Omega^2(v_\phi)+6\xi^2v_\phi^2\bigr)y_i ,
  \label{eq:reheaty}
\end{align}
where small $\bar\chi$ contributions are neglected. 
Furthermore, the decay of RH neutrinos into the SM sector is
sufficiently effective, unlike in the $u_\phi=0$ case.  
The scattering contribution is subdominant, as can be seen from
\eqref{eq:TD} and \eqref{eq:TS}, and will be omitted in the analysis
below. The reheating temperature is then given by \eqref{eq:TN} with
$g_*=112$ if it is smaller than $T_D$ and hence the
decay process is effective until the universe is heated up to
\eqref{eq:TN}, leading to the constraint
\begin{align}
  \lambda^{1/2} < 1.5\times10^4\,v_\phi
  \sqrt{\Omega^2(v_\phi)+6\xi^2v_\phi^2} \,
  \Big(\frac{m_{\nu_i}}{10^{-1}\,\text{eV}}\Big)^2 .
  \label{eq:TRTD}
\end{align}
If this condition fails, the reheating temperature for the SM sector
is determined by $T_D$. We find the inequality \eqref{eq:TRTD}
generally holds in the present inflation scenario unless $v_\phi$ is
very small, and the reheating temperature $T_R$ is given by
\eqref{eq:TN} with $g_*=112$, namely,
\begin{align}
   T_R =0.052\, \frac{\lambda^{1/4}y_i v_\phi^{1/2}}
   {(\Omega^2(v_\phi)+6\xi^2v_\phi^2)^{1/4}}
   \biggl(1 - \frac{4(\Omega^2(v_\phi)+6\xi^2v_\phi^2)^2y_i^2}
   {\Omega^4(v_\phi)\lambda}\biggr)^{3/4},
   \label{eq:TR2}
\end{align}
where the phase-space factor is explicitly included, though
subdominant for evaluating $T_R$. 
As seen from Figure~\ref{fig:LGyukawa}, the reheating temperature is not
continuously connected to the $u_{\phi}=0$ case. For small $\xi$, it
scales as $T_R \propto \xi^{1/2}$ if a rough estimation
$\lambda \propto \xi^2$ inferred from the power spectrum is assumed. On
the other hand, for larger $\xi$ ($\xi v_\phi \gg 1$), $T_R$ remains
almost constant and depends linearly only on the inflaton coupling
$y_i$. The minimum requirement for successful reheating,
\eqref{eq:reheaty}, also leads to a non-trivial upper bound on
$y_i$. In Figure~\ref{fig:LGyukawa}, the gray regions are excluded by
this reheating condition. A smaller $y_i$ is favored, as can be seen
from the fact that the blue region in Figure~\ref{fig:LGyukawa} becomes
wider as $y_i$ decreases. For a fixed $y_i$, the condition \eqref{eq:reheaty}
turns out to yield a lower (upper) bound on $\xi$ for small (large)
values of $\xi$, if the rough estimation $\lambda \propto \xi^2$ is
taken into account. This feature indeed corresponds to the horizontal
brown-hatched bands in the $(m_\chi,\xi)$ plane.

The thermal leptogenesis is also possible for the $u_\phi \neq 0$ case.
The reheating temperature must be at least higher than the mass of the
lightest RH neutrino, which implies a parameter constraint derived from
Eqs.~\eqref{Formula:massofN} and \eqref{eq:TR2},
\begin{align}
  4y_i^2 < \frac{\lambda\Omega^4(v_\phi)}
  {(\Omega^2(v_\phi)+6\xi^2v_\phi^2)^2} 
  -\frac{50(\Omega^2(v_\phi)\lambda v_\phi)^{2/3}}
  {\Omega^2(v_\phi)+6\xi^2v_\phi^2}.
  \label{eq:leptoy}
\end{align}
This bound is satisfied in the brown hatched regions of
Figure~\ref{fig:LGyukawa}, where the leptogenesis can successfully take
place. Assuming that $\lambda$ and $\xi$ are determined by the inflation
dynamics and $v_\phi$ is treated as a fixed parameter in our analysis,
the condition \eqref{eq:leptoy} provides an upper bound on the
inflaton Yukawa coupling $y_i$. This represents the phase-space
constraint of the inflaton decay and does not strongly affect the
allowed parameter space once the decay channel is kinematically
open. These behaviors can be seen from the variation of $y_i$ in
Figure~\ref{fig:LGyukawa}. It should be noted that the 
condition \eqref{eq:leptoy} (corresponding to the brown hatched 
region) is necessarily stronger than \eqref{eq:reheaty} (the blue
region), since the latter corresponds to the vanishing limit of the
phase space. For a fixed $y_i$, the condition \eqref{eq:leptoy}
determines the bound on $\xi$ almost independently of
$m_\chi$. Considering again the approximate 
relation $\lambda \propto \xi^2$ inferred from the power 
spectrum, one can derive a lower (upper) bound on $\xi$ for small
(large) values of $\xi$, which corresponds to the horizontal
brown-hatched bands in the $(m_\chi,\xi)$ plane.

Finally, we comment on several other possible reheating mechanisms in
the present inflation model. In the gray region of
Figure~\ref{fig:LGyukawa}, the inflaton decay into heavy $N_i$ is
forbidden, and the main decay mode to the SM sector may be the top-quark
pair channel. The reheating through this mode alone yields an approximate
temperature $T_R \sim 10^{-14}\lambda^{1/4}v_\phi^{1/2}
(m_{\nu_i}/\text{eV})^2$, which is suppressed by the large $N_i$ masses
and typically falls within
$\mathcal{O}(\text{MeV})$–$\mathcal{O}(\text{GeV})$ for the parameters
in the gray regions. Such a
temperature is  just before the big bang nucleosynthesis but would need
to be enhanced, for instance, to achieve successful leptogenesis. As an
alternative possibility, the inflaton may also interact with the scalar
sector through the Higgs-portal coupling, inflaton potential mixings, or
other related terms. While this type of reheating process is beyond
the scope of the present work, it would be interesting to explore it in
connection with the inflaton shift toward the non-trivial vacuum, and
we leave this topic for future investigation.

\medskip

\section{Summary}
\label{sec:conclusions}

We have studied a two-field inflation model originating from a single
complex scalar field that acquires a non-vanishing vacuum expectation
value. The soft-breaking scalar mass gives rise to two inflaton degrees
of freedom: the radial mode and the Nambu–Goldstone mode associated with
the spontaneous symmetry breaking. By introducing an effective e-folding
number, we have classified the inflationary trajectories into three
types according to which mode dominantly drives the inflationary
expansion: the Higgs-inflation type, the natural-inflation type, and
the mixed type. We have shown the model parameter space consistent
with current cosmological observations and found that all three types
are allowed. While the Higgs-type scenario is broadly favored for
large values of the non-minimal coupling to gravity, the coexistence of the
Nambu–Goldstone mode also enables successful inflation even for smaller
non-minimal couplings, leading to either the mixed or the natural type
of inflation. It is particularly notable that, though the natural
inflation is generally known to require a large symmetry-breaking scale,
successful inflation with a smaller symmetry-breaking scale can occur in
the present model due to the contribution from the radial mode with a
large field value and a small non-minimal coupling.

Based on the inflationary dynamics, we have further examined whether
the reheating and leptogenesis can be achieved through the inflaton decay to
right-handed neutrinos, where the same complex scalar field plays
the role of the inflaton and generates the neutrino Majorana
masses. The scalar field can deviate from the potential local minimum
due to a non-trivial field value remaining after the inflationary expansion.
Taking such an effect into account and evaluating the reheating temperature
and leptogenesis, we have identified the viable region of model
parameters that consistently realizes the inflation, the reheating, and
the observed matter–antimatter asymmetry. In addition to the
conventional oscillation around the non-trivial vacuum, as in the
Higgs inflation with a non-minimal coupling, we have found that a
region with a large soft-breaking mass is also allowed, where the
inflaton oscillates around the origin in field space. 
This is one of the characteristic features of the present two-field
model. A more detailed and quantitative investigation, including the
effect of vacuum shift and possible additional couplings of the
inflaton fields, would provide deeper insights into the post-inflationary
dynamics. These aspects will be systematically explored in future work.

\section*{Acknowledgments}
\noindent
The work of Y.A.\ is supported by JSPS Overseas Research Fellowship.
The work of K.Y.\ is supported by JSPS KAKENHI Grant No.~JP20K03949.

\appendix

\medskip

\section{Scalar field dynamics}
\label{app:scalar-sector}

In this appendix, we summarize the detailed formulation of the scalar
sector and its dynamical equations.

\subsection{EOMs}
\label{app:EOM}

The general form of the action for scalar fields $\varphi^a$ is written as
\begin{align}
  S &= \int d^4x \sqrt{-g} \biggl[
      - \frac{R}{2} + \frac{1}{2} K_{ab}(\varphi)
      \del_\mu \varphi^a \del^\mu \varphi^b - V(\varphi)
    \biggr].
\end{align}
From this action, the equation of motion (EOM) for the scalar field is
derived as 
\begin{align}
  \nabla^2 \varphi^a + \gamma^a_{bc}(\varphi)
  \del_\mu \varphi^b \del^\mu \varphi^c
  + K^{ab}(\varphi) \del_b V(\varphi) = 0,
\end{align}
where $\nabla$ denotes the covariant derivative with respect to the
Einstein-frame metric $g_{\mu\nu}$. The Levi-Civita connection in the
field space is defined by 
\begin{align}
  \gamma^a_{bc} = \frac{1}{2} K^{ad}
  (\del_b K_{dc} + \del_c K_{bd} - \del_d K_{bc} ).
\end{align}
We consider the spatially flat Friedmann–Lemaître–Robertson–Walker metric,
\begin{align}
  ds^2 = g_{\mu\nu} dx^\mu dx^\nu = dt^2 - a(t)^2 d \vec{x}^2,
\end{align}
and assume that $\varphi^a$ depends only on cosmic time $t$.
Under these assumptions, the general EOM reduces to
\begin{align}
  \ddot{\varphi}^a + 3 H \dot{\varphi}^a
  + \gamma^a_{bc} \dot{\varphi}^b \dot{\varphi}^c + K^{ab} \del_b V = 0,
  \label{eq:EOM-t}
\end{align}
where a dot denotes the derivative with respect to $t$, and
$H \coloneqq \dot{a}/a$ is the Hubble parameter.
By introducing the e-folding number $N$ via $dN = H dt$,
Eq.~\eqref{eq:EOM-t} can be rewritten as
\begin{align}
  \frac{d^2 \varphi^a}{d N^2} + \gamma^a_{bc} \frac{d \varphi^b}{dN}
  \frac{d \varphi^c}{dN} + (3 -\varepsilon)
  \biggl( \frac{d \varphi^a}{dN} + K^{ab} \del_b \ln V \biggr) =0.
\end{align}
In this derivation, we have used the Friedmann equation, 
\begin{align}
  H^2 = \frac{1}{3} \biggl(
  \frac{1}{2} K_{ab} \dot{\varphi}^a \dot{\varphi}^b
  + V(\varphi) \biggr),
  \qquad 
  \dot{H} = - \frac{1}{2} K_{ab} \dot{\varphi}^a \dot{\varphi}^b,
\end{align}
and introduced the slow-roll parameter $\varepsilon$, defined by 
\begin{align}
  \varepsilon \coloneqq - \frac{\dot{H}}{H^2}
  = \frac{1}{2} K_{ab} \frac{d \varphi^a}{dN} \frac{d\varphi^b}{dN}.
\end{align}
This expression is useful for describing multi-field slow-roll
dynamics in terms of field-space quantities.

\subsection{Geometry of scalar fields}

We next consider the case of a single complex scalar field,
$\Phi=\phi e^{i\chi}/\sqrt{2}$, with a non-minimal coupling $\xi$, as
introduced in Section~2. The field-space metric is then found to be
\begin{align}
  K_{ab} = \diag \left(
  \frac{1 + \xi \phi^2 + 6 \xi^2 \phi^2}
  {(1 + \xi \phi^2)^2},\, \frac{\phi^2}{1 + \xi \phi^2}
  \right).
\end{align}
The non-vanishing components of the Levi-Civita connection in this
field space are given by
\begin{align}
  \gamma^1_{11} &= - \frac{\xi \phi ( 1 -6 \xi
    +\xi \phi^2 +6 \xi^2 \phi^2 )}
    {(1 + \xi \phi^2)(1 + \xi \phi^2 + 6 \xi^2 \phi^2)}, 
  \\
  \gamma^1_{22} &= - \frac{\phi}{1 + \xi \phi^2 + 6 \xi^2 \phi^2}, 
  \\
  \gamma^2_{12} &= \gamma^2_{21} = \frac{1}{\phi(1 + \xi \phi^2)}. 
\end{align}
Here the indices $1$ and $2$ correspond to the fields
$\phi$ and $\chi$, respectively.
The Ricci scalar of the field-space geometry is then obtained as 
\begin{align}
  \msc{R} = \frac{4 \xi ( 1 + 3\xi + \xi\phi^2 + 6 \xi^2 \phi^2 )}
  {(1 + \xi \phi^2 + 6 \xi^2 \phi^2)^2}.
  \label{Formula:RicciScalar}
\end{align}
It is worth noting that the curvature vanishes in the limit
$\xi \to 0$, where the field space becomes flat.

\medskip

\section{Slow-roll and slow-turn parameters}
\label{app:SRparameters}

In this appendix, we present the definitions and expressions for the
slow-roll and slow-turn parameters that characterize the multi-field
inflation dynamics. These quantities describe, respectively, the
magnitude of the field velocity and the curvature of its trajectory in
field space.

In the field space spanned by $\varphi^a$, the field velocity is given
by its derivative with respect to the e-folding number,
$d\varphi^a/dN$, while the (covariant) acceleration of the field
vector is defined as
$\eta^a=d^2\varphi^a/dN^2+\gamma^a_{bc}(d\varphi^c/dN)(d\varphi^b/dN)$.
The slow-roll and slow-turn parameters are expressed in terms of the
norm of the velocity vector and the parallel and perpendicular
components of the acceleration vector as~\cite{Peterson:2010np}:
\begin{align}
  \varepsilon &= 
  \frac{1}{2}K_{ab}\frac{d\varphi^a}{dN}\frac{d\varphi^b}{dN},
  \\
  \frac{\eta_\parallel^a}{v} &= 
  \frac{1}{v}\, K_{ab}\hat{e}_\parallel^a \Big(\frac{d^2\varphi^b}{dN^2} 
  +\gamma^b_{dc}\frac{d\varphi^c}{dN}\frac{d\varphi^d}{dN}\Big), 
  \\
  \frac{\eta_\perp^a}{v} &= \frac{1}{v}\, K_{ab}\hat{e}_\perp^a
    \Big(\frac{d^2\varphi^b}{dN^2} 
    +\gamma^b_{dc}\frac{d\varphi^c}{dN}\frac{d\varphi^d}{dN}\Big).
\end{align}
For the acceleration vector, dividing by the speed $v$ expresses the
rate of change of the field velocity relative to the expansion rate.
When the slow-roll and slow-turn approximations hold,
$\varepsilon,\, |\eta_\parallel|/v,\, \eta_\perp/v \ll 1$, these
parameters can be explicitly written in terms of the inflaton
potential and the field-space metric, as given in
Eqs.~\eqref{formula:epsilon}--\eqref{formula:etaPerpov}.

We now present the explicit forms of the slow-roll and slow-turn
parameters for our two-field inflation model, using the approximations
employed in the main text.
When the soft-breaking mass $m_{\chi}$ is sufficiently large, the
inflationary dynamics are dominated by the $\chi$ potential. In this
regime, the slow-roll parameters are given by
\begin{align}
  \varepsilon &\approx \frac{2}{\phi^2} \bigg[
    \frac{ (1 -\xi \phi^2)^2}{1 + \xi \phi^2 + 6 \xi^2 \phi^2}+ (1+\xi \phi^2) \cot^2\chi\bigg],
  \\
  \frac{\eta_\parallel}{v} & \approx \frac{ 1 + \xi \phi^2 }{\phi^2}
  \Bigl[\, \frac{2(1 - \xi \phi^2)^2 ( 1 + 3 \xi \phi^2 + 12 \xi^2 \phi^2) \tan^2\chi}{(1 + \xi \phi^2 + 6 \xi^2 \phi^2 )^2}  \nn \\
  & \quad +\csc^2 \chi \bigl[
    1 + 5 \xi \phi^2 + 12 \xi^2 \phi^2 + 2 \xi^2 \phi^4 + 12 \xi^3 \phi^4 
    + (1 - \xi\phi^2)\cos2\chi	\bigr]   \nn \\
  & \qquad + 4(1- \xi \phi^2) \Big] \Big/ \Big[
    ( 1 -\xi \phi^2)^2 \tan^2 \chi + (1 + \xi \phi^2)(1 + 
    \xi \phi^2 + 6 \xi^2 \phi^2) \Big].
\end{align}
Both parameters are independent of $m_{\chi}$ because the $\chi$
potential is directly proportional to $m_\chi^2$.

When $\phi$ takes a value larger than $v_\phi$ during inflation, the
slow-roll parameters can be approximated as 
\begin{align}
  \varepsilon & \approx \frac{8}{\lambda^{2} \xi ( 1 + 6 \xi )\phi^4}
    \Big[ \big[\lambda(1 +\xi v_\phi^2)
    - 2\xi m_\chi^2 \sin^2 \chi\big]^2
    + m_\chi^4 \xi^2 ( 1 + 6 \xi)\sin^2 2\chi \Big], \\
  \frac{\eta_\parallel}{v} & \approx \frac{8 \xi}{\lambda \phi^2}
    \biggl[
	\frac{\big[\lambda(1 +\xi v_\phi^2)
        - 2\xi m_\chi^2 \sin^2 \chi\big]^3}
        {\xi^2 ( 1 + 6 \xi )^2} + \frac{2\big[\lambda(1 +\xi v_\phi^2)
        - 2\xi m_\chi^2 \sin^2 \chi\big] m_\chi^4
        \sin^2 2\chi}{1 + 6 \xi}   \nn \\
  & \qquad  - \xi m_\chi^6 \cos 2\chi \sin^2 2\chi  \biggr] 
    \bigg/  \biggl[
  \frac{\big[\lambda(1 +\xi v_\phi^2) - 2\xi m_\chi^2 \sin^2 \chi\big]^2}
    {\xi ( 1 + 6 \xi)} + \xi m_\chi^4 \sin^2 2\chi 
    \biggr] ,
\end{align}
where the dependence on both $\xi$ and $m_\chi$ determines the
deviation from the simple single-field slow-roll limit.
On the other hand, in the small-field region where $\phi < v_\phi$,
the slow-roll parameters take the form
\begin{align}
  \varepsilon & \approx \frac{8 \phi^2}{\lambda^2 v_\phi^8} \Big[
    \big[\lambda v_\phi^2(1 + \xi v_\phi^2)
     - 2m_\chi^2 \sin^2\chi\big]^2
     + m_\chi^4 \sin^2 2\chi \Big], \\
  \frac{\eta_\parallel}{v} & \approx \frac{4}{\lambda v_\phi^4} 
     \Big[\big[\lambda v_\phi^2(1 + \xi v_\phi^2)
     - 2m_\chi^2 \sin^2\chi\big]^3 + 3\big[\lambda v_\phi^2(1 + \xi v_\phi^2)
     - 2m_\chi^2 \sin^2\chi\big] m_\chi^4  \sin^2 2\chi \nn \\
 & \qquad  - 2 m_\chi^6 \cos 2\chi \sin^2 2\chi  \Big] \Big/
   \Big[ \big[\lambda v_\phi^2(1 + \xi v_\phi^2)
   - 2m_\chi^2 \sin^2\chi\big]^2 + m_\chi^4 \sin^2 2\chi \Big].
\end{align}
These expressions illustrate how the soft-breaking mass and the
non-minimal coupling jointly determine the inflationary dynamics in
the small-field regime.

\medskip

\section{Formulae of cosmological observables}
\label{app:observables}

\subsection{Transfer functions}

In multi-field inflation, isocurvature perturbations can convert into
curvature perturbations after horizon exit. Therefore, it is crucial to
formulate the transfer functions that describe the super-horizon
evolution of these coupled modes. Following
Ref.~\cite{Peterson:2010np}, we define the effective mass matrix as
\begin{align}
  M^{a}{}_{b} \coloneqq 
  K^{ac} (\del_c \del_b \ln V-\gamma^d_{cb}\del_d\ln V) 
  + \frac{1}{3} \varepsilon \msc{R} \hat{e}_{\perp}^{a} K_{bc}\hat{e}_\perp^c.
  \label{eq:Mtilde}
\end{align}
Given the expressions for the basis vectors
\eqref{formula:eperp} and the Ricci scalar in the field space
\eqref{Formula:RicciScalar}, the matrix $M$ is completely determined by
the model parameters and the background field values during inflation.
Using this matrix, the transfer functions are defined as
\begin{align}
  T_{\mc{SS}} &= \exp\bigg[ \int_{N_*}^{N} dN'
  (K_{ac}\hat{e}_\parallel^c M^{a}{}_{b} \hat{e}_{\parallel}^{b}
  -K_{ac} \hat{e}_\perp^c M^{a}{}_{b} \hat{e}_{\perp}^{b} ) \bigg], 
  \label{eq:TSS}
  \\
  T_{\mc{RS}} &= \int_{N_*}^{N} dN' \,\frac{ \eta_{\perp} }{ v }
  \exp\bigg[ \int_{N_*}^{N'} dN''
  (K_{ac}\hat{e}_\parallel^c M^{a}{}_{b} \hat{e}_{\parallel}^{b}
  - K_{ac}\hat{e}_\perp^c M^{a}{}_{b} \hat{e}_{\perp}^{b} )\bigg].
  \label{eq:TRS}
\end{align}
These transfer functions depend on the couplings and field
values through the basis vectors and the slow-roll parameter
\eqref{formula:etaPerpov}. Both $T_{\mc{SS}}$ and $T_{\mc{RS}}$ are
positive definite, and they describe how the curvature and isocurvature
modes evolve and mix on super-horizon scales.

\subsection{Cosmological observables}

We now present explicit expressions for the cosmological observables in
the present two-field model. The power spectrum and the tensor-to-scalar
ratio are given as
\begin{align}
  \cPR &= \frac{ (1 + \xi \phis^{2})^4 ( 1 + T_{\mc{RS}}^{2})
    V( \phis, \chis )^3 }{ 3 \pi^{2} \phis^{2}  }   \nn \\
  &\qquad \bigg/ \bigg[
    \frac{[\lambda ( 1 + \xi v_\phi^2)(\phis^2 - v_\phi^2)
    + 2m_\chi^2 (1 -\xi \phis^2)\sin^2 \chis]^{2}}
    {1 + \xi \phis^2 + 6 \xi^2 \phis^2}
    + m_\chi^4 ( 1 + \xi \phis^2 ) \sin^2 2\chis
  \bigg],  \\
  r &= \frac{2 \phis^{2}}{(1 + \xi \phis^{2})^4(1 + \TRS^2)
     V(\phis,\chis)^{2} } \nn  \\
  &\qquad \times \bigg[
    \frac{[\lambda ( 1 + \xi v_\phi^2)(\phis^2 - v_\phi^2)
    + 2m_\chi^2 (1 -\xi \phis^2)\sin^2\chis]^{2}}
    {1 + \xi \phis^2 + 6 \xi^2 \phis^2}
    + m_\chi^4 ( 1 + \xi \phis^2 ) \sin^2 2\chis
   \bigg].
\end{align}
The scalar power spectrum, determined by the correlation function of
scalar fluctuations, grows with increasing $T_{\mc{RS}}$, whereas the
tensor-to-scalar ratio $r$ scales inversely with $(1+T_{\mc{RS}}^2)$.
Hence, the presence of isocurvature-to-curvature transfer tends to
enhance $\cPR$ and suppress $r$.

The spectral index $n_s$ and its running $\alpha$ can also be
expressed in terms of the background field values under the slow-roll
and slow-turn approximation~\cite{Peterson:2010np}:
\begin{align}
  n_{s} &= 1 - 2 \varepsilon_{*}
    + 2 K_{ab*}\hat{e}_{N *}^a M^b{}_{c *} 
    \hat{e}_{N*}^c , \\ 
  \alpha &= [-K^{ab}\partial_a \ln V
    \partial_b (- 2 \varepsilon
    + 2 K_{cd}\hat{e}_N^c M^d{}_e \hat{e}_{N}^e )]_* ,
\end{align}
where the unit vector along the final adiabatic direction is
\begin{align}
  \hat{\bm{e}}_{N} = \frac{1}{\sqrt{ 1 + T_{\mc{RS}}^{2} }}
  \hat{\bm{e}}_{\parallel}
  + \frac{ T_{\mc{RS}} }{ \sqrt{ 1 + T_{\mc{RS}}^{2} }} \hat{\bm{e}}_{\perp}.
\end{align}
Finally, the isocurvature fraction is given by
\begin{align}
  \beta_\iso = \frac{\TSS^2}{1 + \TRS^2 + \TSS^2}.
\end{align}
By substituting the explicit expressions for the basis vectors
\eqref{formula:eparallel} and \eqref{formula:eperp}, as well as the
transfer functions \eqref{eq:TSS} and \eqref{eq:TRS} defined using the
mass matrix \eqref{eq:Mtilde}, all the cosmological observables can be
expressed directly in terms of the parameters and field values of the
present inflation model. Various approximations of these formulae,
appropriate to specific parameter regimes, are discussed in the text.

\medskip

\section{Inflation with various field values}
\label{sec:inflation-patterns}

In this appendix, we consider three patterns of inflation in addition
to those discussed in Section~\ref{Sec:Constraints}. These patterns lead
to different inflaton trajectories and allowed parameter spaces,
depending on the initial field values and the magnitude of the symmetry
breaking scale:
\begin{itemize}
\item $\phi_* > v_\phi = \mc{O}(10)$
\item $\phi_* < v_\phi = \mc{O}(10)$
\item $\phi_* > v_\phi = \mc{O}(1)$
\end{itemize}
We impose the e-folding condition $N = 55$ throughout this appendix.

\subsection{$\phis > v_\phi$ and $v_\phi = \mc{O}(10)$}

\begin{figure}[t]
  \centering
  \begin{subfigure}{0.32\textwidth}
    \centering
    \includegraphics[width=\textwidth]{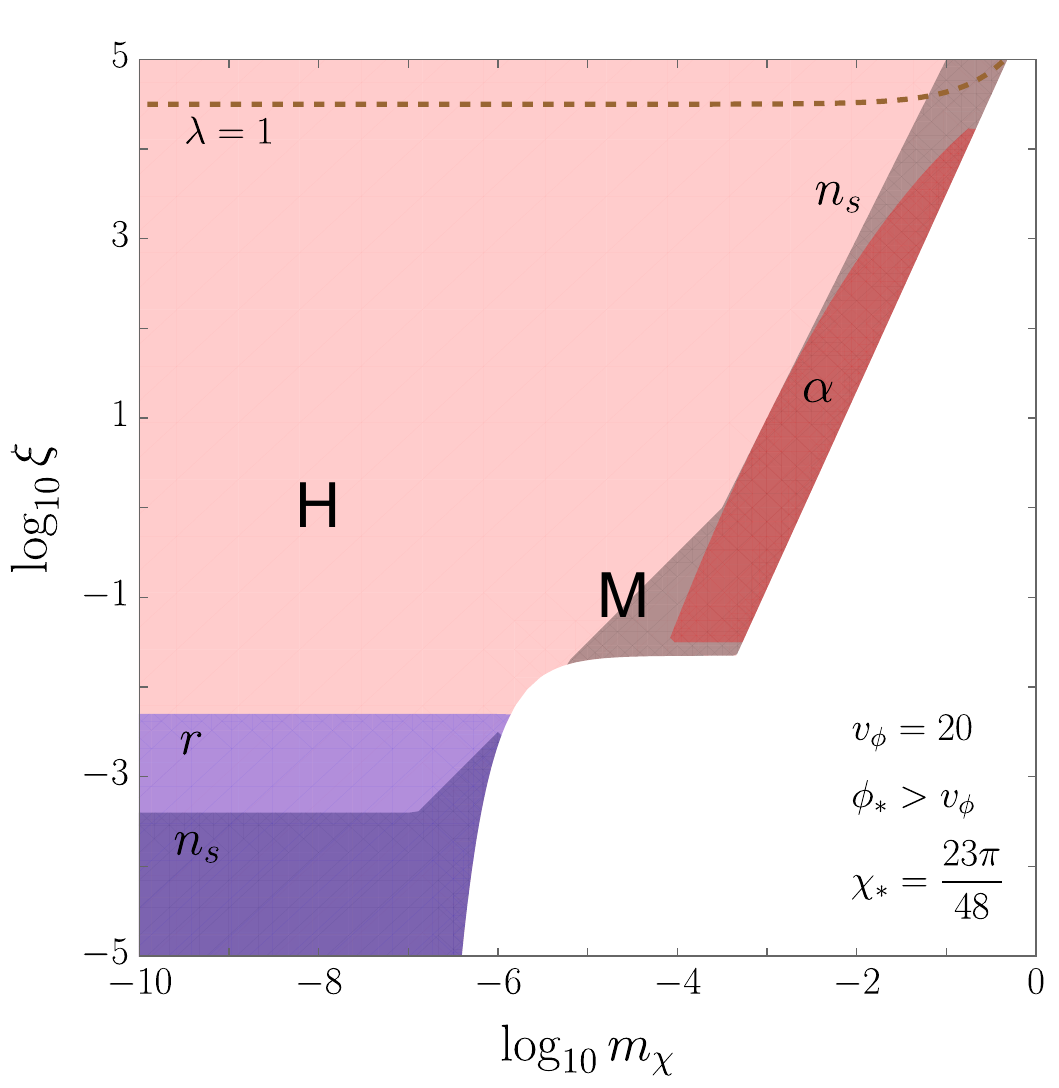}
    \caption{$\chis = 23\pi/48$}
  \end{subfigure}\ 
  \begin{subfigure}{0.32\textwidth}
    \centering
    \includegraphics[width=\textwidth]{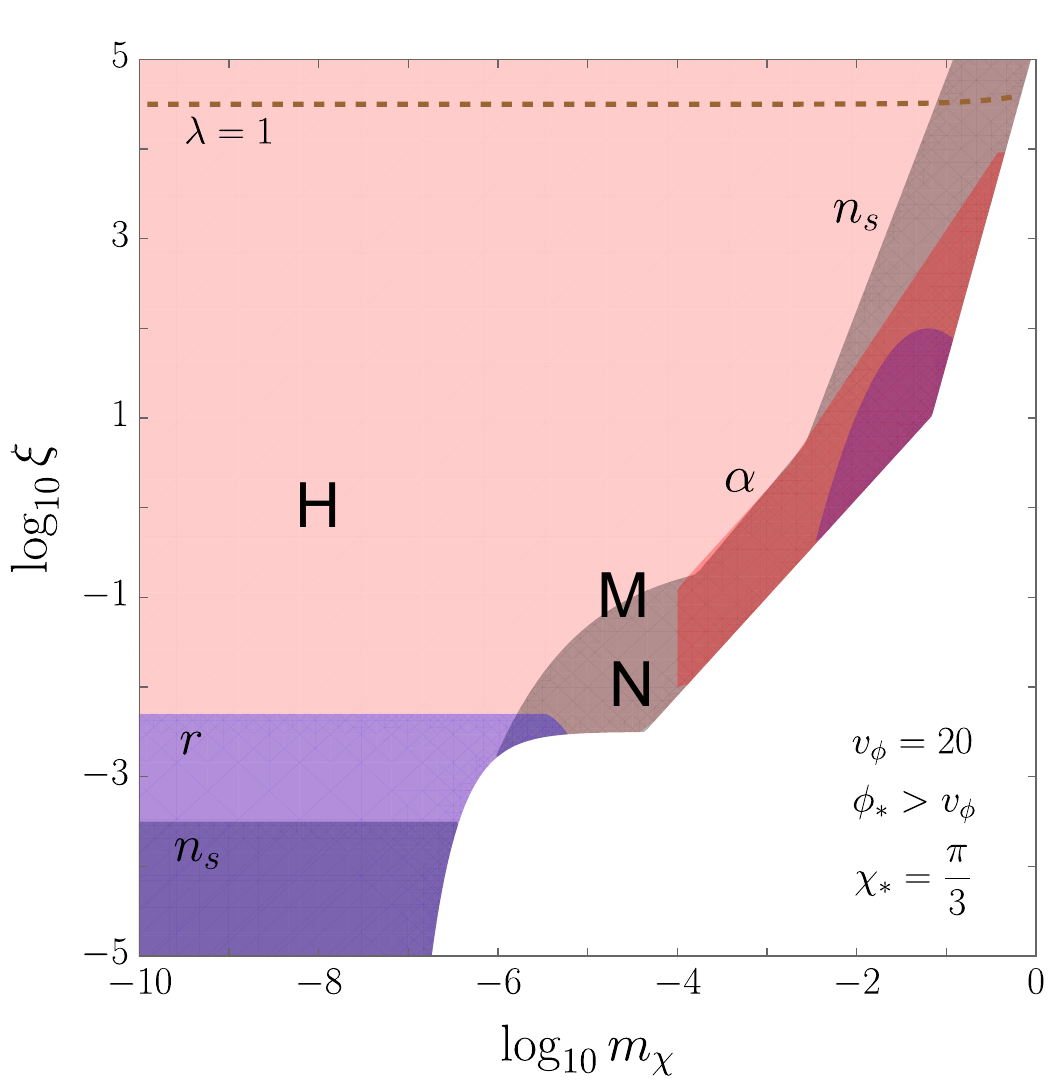}
    \caption{$\chis = \pi/3$}
  \end{subfigure}\ 
  \begin{subfigure}{0.32\textwidth}
    \centering
    \includegraphics[width=\textwidth]{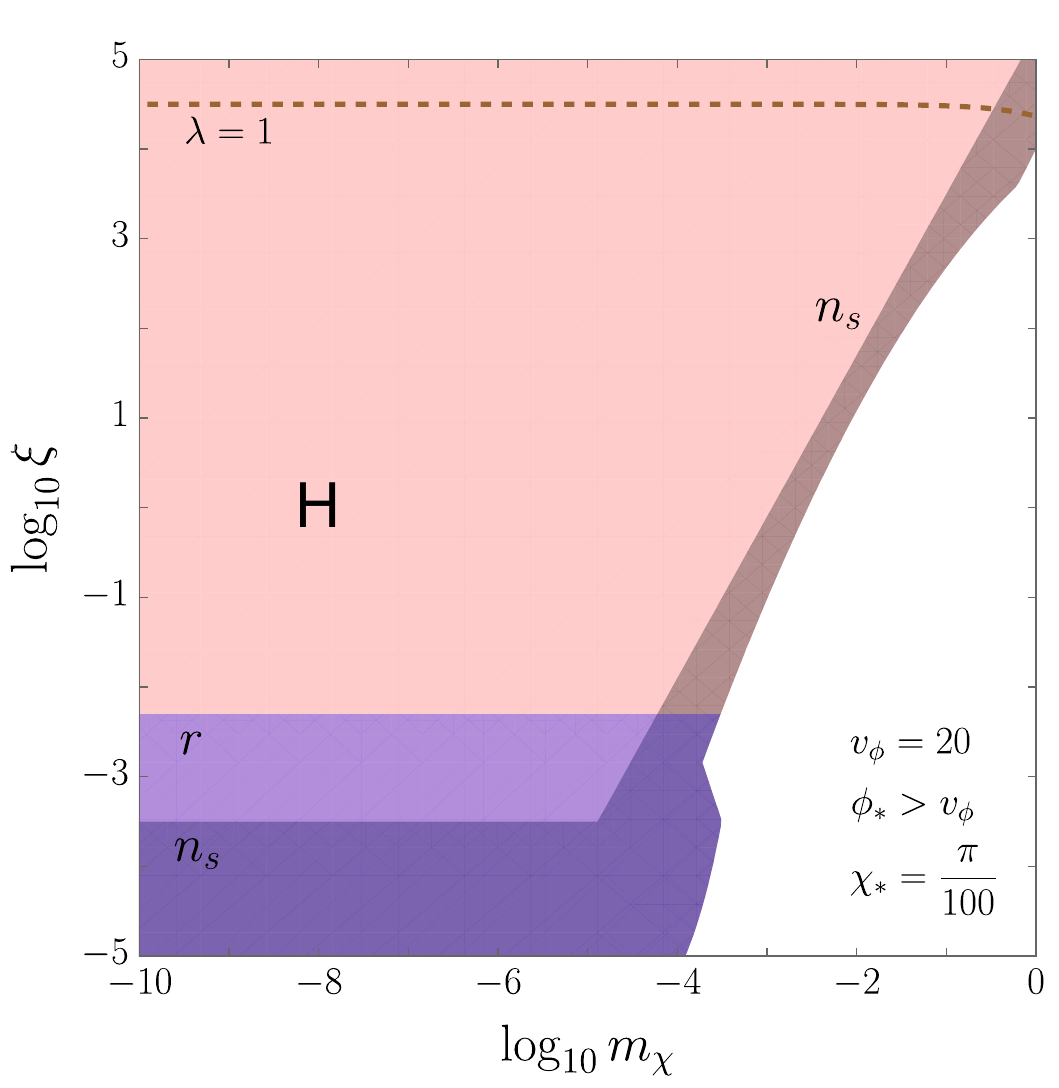}
    \caption{$\chis = \pi /100$}
  \end{subfigure}
  \caption{
    The allowed parameter region in the $(m_\chi, \xi)$ plane for
    $\phis > v_\phi = 20$. The pink region is consistent with the
    cosmological observations and $N=55$.
    The gray, blue, and red regions are excluded by the constraints
    from observations on $n_s$, $r$, and $\alpha$, respectively.
    The scalar quartic coupling $\lambda = 1$ holds along the brown
    dashed lines.}
  \label{Figure:mchixiVEV20}
  \bigskip
\end{figure}

First, we consider $\phis > v_\phi$ with the VEV taking a relatively
large value, $v_\phi = 20$.
Figure~\ref{Figure:mchixiVEV20} shows the allowed parameter region for
inflation under this field value and symmetry-breaking scale.
In the white region, the e-folding condition $N = 55$ cannot be achieved
for large $m_\chi$. We also find that the mixed-type inflation can be
realized around $m_\chi \sim 10^{-5}$ when the VEV is larger than $\mc{O}(1)$.
In this mixed type, the inflaton components that start and end inflation
can be different, corresponding to the so-called hybrid inflation
scenario~\cite{Linde:1993cn}.

\paragraph{\underline{Small $m_\chi$ region}:}

In the small $m_\chi$ region with $\phis > v_\phi = 20$ and $\xi \gg 1$,
the spectral index, the tensor-to-scalar ratio, and running of the spectral
index are approximately given by 
\begin{align}
  n_s & \approx 1 - \frac{ 8 v_{\phi}^2 }{ 3 \phis^2 },
  \label{Formula:Approns2}
  \\
  r & \approx \frac{ 64 v_{\phi}^4 }{ 3 \phis^4 },
  \label{Formula:Appror2}
  \\
  \alpha& \approx - \frac{ 32 v_{\phi}^4 }{ 9 \phis^4 }.
  \label{Formula:Approalpha2}
\end{align}
The relation between $\phis$ and the e-folding number is
\begin{align}
  N = \frac{(1 + 6 \xi)(\phis^2- \phi_e^2)
  - v_\phi^2 \ln \frac{\phis^2}{\phi_e^2} - 6 ( 1 + \xi v_\phi^2)
  \ln \frac{1 + \xi \phis^2}{1 + \xi \phi_e^2}}{8 ( 1 + \xi v_\phi^2)},
  \label{Formula:ApproN2}
\end{align}
where $\phi_e$ denotes the field value at the end of inflation, obtained
from the slow-roll and slow-turn conditions. From these
approximations, we find that inflation in this case is 
consistent with the observational data, mainly because a large $\xi$
flattens the inflaton potential. 
For small $\xi$, however, the potential becomes steeper, leading to
inconsistency between theoretical predictions and observations, as seen
in Figure~\ref{Figure:mchixiVEV20}. Since $\Tilde{\beta}$ is
approximately given by \eqref{Formula:beta1} and
\eqref{Formula:beta11}, the resulting $\beta_{\mathrm{iso}}$ is
sufficiently small. The inflaton behavior discussed here is similar to
that in the case with a small VEV $v_\phi \ll 1$, because the
inflationary motion is mainly driven by the radial mode.

\paragraph{\underline{Large $m_\chi$ region}:}

In this case, the inflaton trajectory strongly depends on the initial
value of $\chi$. For large $m_\chi$, the pNGB 
potential gives a significant contribution, allowing the inflaton to
move along the $\chi$ direction. In particular, the 
natural inflation can be realized if the initial field values and
parameters are appropriately tuned, as seen in the middle panel of
Figure~\ref{Figure:mchixiVEV20}. 
When both scalar fields participate in the inflationary dynamics, a
larger $\xi$ leads to results consistent with observations, as shown in
the left and middle panels. 
In contrast, small $\xi$ results in the chaotic or natural inflation, both
of which are excluded by observational constraints.

For large symmetry breaking, there exists an upper bound on the soft
breaking mass $m_\chi$. 
For $\xi \gg 1$, the power spectrum can be approximated as 
\begin{align}
  \mc{P_{R}} \approx
  \frac{ (\lambda \phis^{2} + 4m_{\chi}^{2} \sin^2\chis )^3
  (1 + T_{\mc{RS}}^{2}) }{ 1536 \pi^{2} \xi^3 \phis^2 m_{\chi}^{4} \sin^2 2\chis}.
  \label{Formula:PRappromchixi1}
\end{align}
A larger $v_{\phi}$ requires a larger $\phis$, which in turn implies a
larger $m_{\chi}$ from the Planck normalization of $\mc{P_R}$. 
Consequently, the experimentally allowed values of $m_{\chi}$ become
larger than those for $v_{\phi} \ll 1$ with $\xi \gg 1$. 
For $\xi \ll 1$, on the other hand, the power spectrum is approximated
as
\begin{align}
  \mc{P_{R}} \approx \frac{ \phis^4 (\lambda \phis^{2}
  + 4m_{\chi}^{2} \sin^2\chis )^3 (1 + T_{\mc{RS}}^{2})}{ 1536 \pi^{2} [ \lambda^2 \phis^4 +4 m_{\chi}^{2} ( \lambda \phis^{2} + m_{\chi}^{2} ) \sin^2 \chis ] }.
\end{align}
When $m_\chi$ is large, the pNGB potential dominates and we find
$\mc{P_{R}} \propto m_{\chi}^{2}$, since the soft-breaking mass
parameter sets the overall coefficient of the inflaton potential. 
Therefore, a larger $v_{\phi}$ (and thus a larger $\phis$) requires a
smaller $m_{\chi}$ for $\xi \ll 1$. 
This behavior is opposite to the $\xi \gg 1$ case.

Regarding the isocurvature bound, the function $\Tilde{\beta}$ behaves
as
\begin{align}
  \Tilde{\beta} &\approx - \frac{2}{\phi^{2}\sin^2\chi}
  \quad (\xi \ll 1), 
  \\
  \Tilde{\beta} &\approx - \frac{2\xi}{\sin^2 \chi}
  \quad (\xi \gg 1),
\end{align}
for $v_\phi \gg 1$ and $m_{\chi}\gg \sqrt{\lambda}\,v_{\phi}$.
Hence, $\beta_\iso$ tends to be exponentially small in this type of
inflation, similar to the case with $v_{\phi} \ll 1$.

\subsection{$\phis < v_\phi$ and $v_\phi = \mc{O}(10)$}

We next discuss the case of inflation with $\phis < v_\phi$, in which
the field $\phi$ rolls down toward the vacuum from a smaller initial
value. We take the symmetry-breaking scale as $v_\phi = 20$. 
For small field values, the slow-roll condition is not satisfied for 
smaller VEV, since $|\eta_\parallel/v|\approx 4/v_\phi^2$.

\begin{figure}[t]
  \centering
  \begin{subfigure}{0.32\textwidth}
    \centering
    \includegraphics[width=\textwidth]{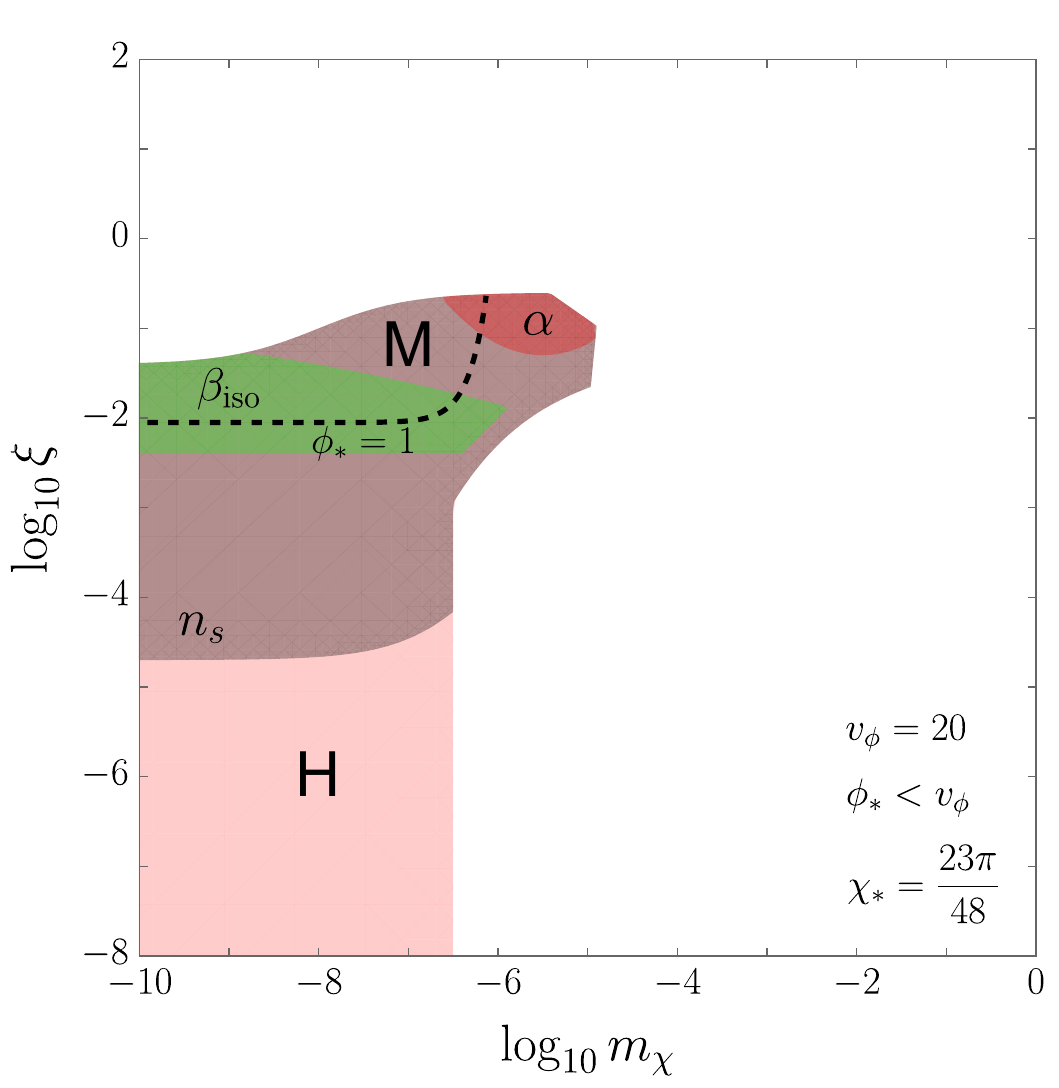}
    \caption{$\chis = 23\pi/48$}
  \end{subfigure}\ 
  \begin{subfigure}{0.32\textwidth}
    \centering
    \includegraphics[width=\textwidth]{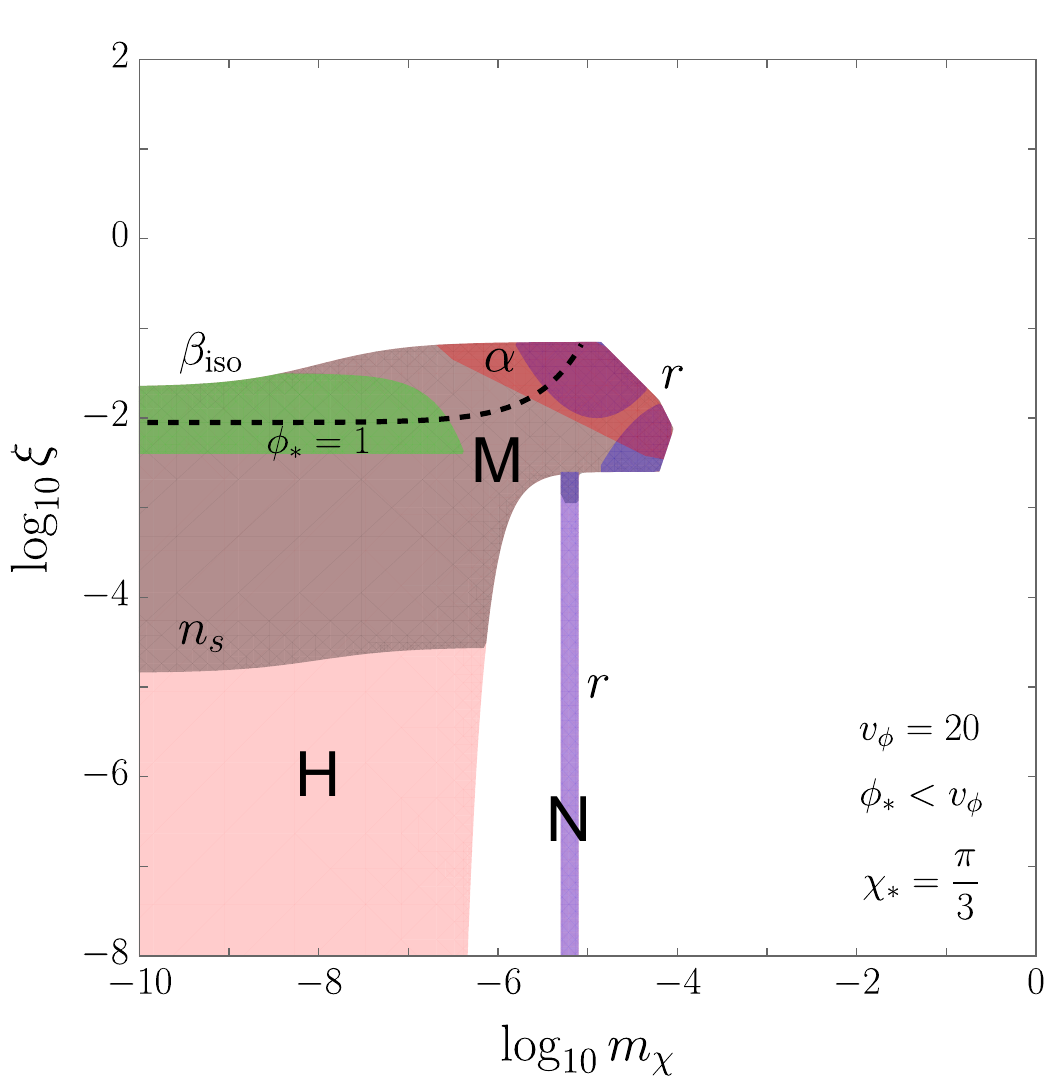}
    \caption{$\chis = \pi /3$}
  \end{subfigure}
  \begin{subfigure}{0.32\textwidth}
    \centering
    \includegraphics[width=\textwidth]{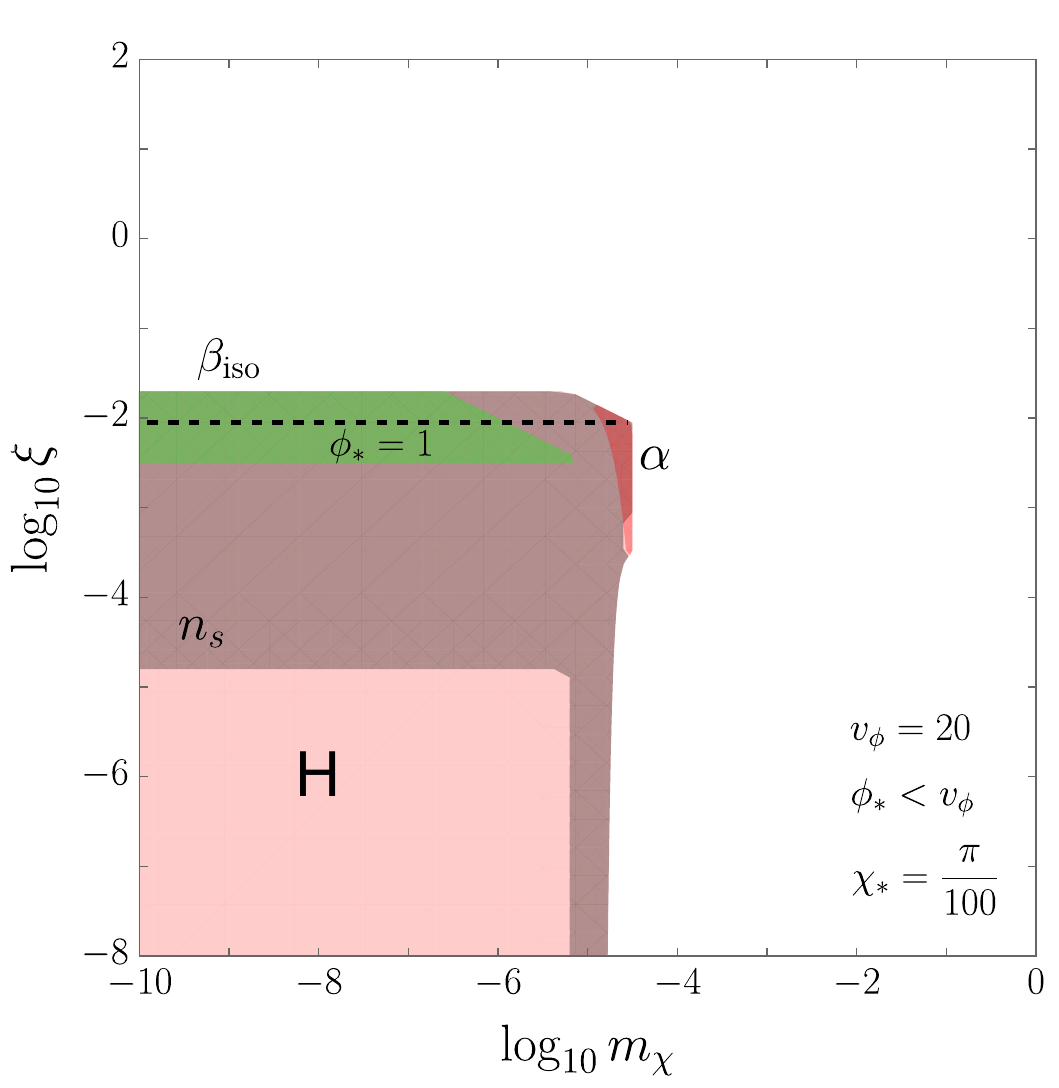}
    \caption{$\chis = \pi/100$}
  \end{subfigure}
  \caption{
    The allowed parameter region in the $(m_\chi, \xi)$ plane for
    $\phis < v_\phi = 20$. 
    The green shaded region is excluded by the isocurvature bound
    $\beta_{\mathrm{iso}}$. 
    The meanings of the other colored regions are the same as in
    Figure~\ref{Figure:mchixiVEV20}. 
    The initial field value $\phi_*=1$ corresponds to the black
    dashed lines.}
  \label{Figure:mchixiVEV20L}
  \bigskip
\end{figure}

Figure~\ref{Figure:mchixiVEV20L} shows the allowed parameter region in
the $(m_\chi, \xi)$ plane for $\phis < v_\phi$ with $N = 55$.
A relatively small $\xi$ is not excluded because the inflaton potential
is sufficiently flat when $\phis < v_\phi$. 
On the other hand, the region with large $\xi$ is excluded, as the potential
becomes too steep for $\phis < v_\phi$. These behaviors are in
contrast to those found in the previous case with $\phis > v_\phi$. 

\paragraph{\underline{Small $m_\chi$ region}:}

When $m_\chi$ is small, the cosmological observables can be expressed as
\begin{align}
  n_s &\approx 1 - \frac{ 8 ( 1 + \xi v_{\phi}^{2} ) }{ v_{\phi}^{2} }
    - \frac{ 8 ( 5  + 6 \xi v_{\phi}^{2} +\xi^2 v_\phi^4 )\phis^{2} }{v_\phi^4},
  \label{Formula:Approns3}
  \\
  r & \approx \frac{ 128 ( 1 + \xi v_{\phi}^{2} )^2 \phis^{2} }{ v_{\phi}^{4} (1+\xi \phis^2)}, 
  \label{Formula:Appror3}
  \\
  \alpha & \approx
   -\frac{64(1+\xi v_\phi^2)^2(5+\xi v_\phi^2-12\xi^2 v_\phi^2)\phis^2}{v_\phi^6}, 
  \label{Formula:Approalpha3}
\end{align}
within the region where the slow-roll and slow-turn conditions hold.

When $\xi\ll1$, both $r$ and $\alpha$ are sufficiently small to be
consistent with observations, due to the suppression by the large symmetry
breaking scale. Moreover, $n_s$ lies within the experimentally allowed
range, as shown in Figure~\ref{Figure:mchixiVEV20L}. 
This behavior again arises from the flat inflaton potential in the
region $\phi_{*}<v_{\phi}$. 

On the other hand, for $\xi \lesssim \mc{O}(1)$, the inflaton potential
becomes more tilted in the region $\phis < v_{\phi}$, leading to a
smaller value of $n_s$ than observed. 
This can be seen from the approximate expression
\begin{align}
  n_s \approx 1 - \frac{8}{v_\phi^2} - 8\xi .
\end{align}
This formula, together with Figure~\ref{Figure:mchixiVEV20L}, shows that
a smaller $\xi$ is required for successful inflation under the present
conditions. 

Finally, we discuss the isocurvature fraction $\beta_{\mathrm{iso}}$. 
The green shaded region in Figure~\ref{Figure:mchixiVEV20L} is excluded
by the observational isocurvature bound. 
In the region with $m_{\chi} \ll v_{\phi}$, the function
$\tilde{\beta}$ in \eqref{Formula:betaisoAppro} can be approximated as
\begin{align}
  \tilde{\beta} \approx
  -\frac{8\phi^{2}}{3v_{\phi}^{4}}(1+\xi v_\phi^2)
  (3 + 4\xi +12\xi^2 -5\xi^2 v_\phi^2 +12 \xi^3 v_\phi^2). 
  \label{Formula:beta2}
\end{align}
For sufficiently small or large $\xi$, $\Tilde{\beta}$ is negative, so
$\beta_{\mathrm{iso}}$ remains small. 
For $\xi \sim \mathcal{O}(1)$, however, $\Tilde{\beta}$ becomes
positive, leading to a large $\beta_{\mathrm{iso}}$ and hence
experimental exclusion, as seen in Figure~\ref{Figure:mchixiVEV20L}.

\paragraph{\underline{Large $m_\chi$ region}:}

In this region, we find three distinct patterns of inflation depending
on the initial value of $\chi$. 
For small $\chis$, inflation is mainly driven by $\phi$, whereas for
large $\chis$, both $\phi$ and $\chi$ contribute, as shown in the left
and middle panels of Figure~\ref{Figure:mchixiVEV20L}. 
These two cases are labeled as H and M, respectively, in the figures. 
As discussed earlier, large values of $\xi$ and $m_\chi$ lead to a small
$n_s$ because the potential of the radial component is dominated by the
quadratic term. Thus, inflation involving both components is difficult
to realize for $\phis < v_\phi$, in contrast to the $\phis > v_\phi$ scenario.

A characteristic feature appears in the region with large $m_{\chi}$ and
small $\xi$, as seen in the middle panel of Figure~\ref{Figure:mchixiVEV20L}. 
Here, inflation similar to the natural inflation occurs for a
specific value of $m_{\chi}$, as previously discussed in
Figure~\ref{Figure:nsr}. 
This arises when the initial value $\chi_{*}$ is tuned so that the
required e-folding $N$ is generated almost entirely by the $\chi$ motion. 
Consequently, $r$ becomes larger than the observed value, consistent
with the approximate relation~\cite{Rubio:2018ogq}
\begin{align}
  r \approx \frac{32}{ \phis^2 \sin^2\chis }.
\end{align}
Since this natural-type inflation is effectively driven by a single
field $\chi$, the typical hierarchy between $|1-n_{s}|$ and $|\alpha|$
is preserved. In this region, where the natural-type inflation occurs,
$\Tilde{\beta}$ can be approximated as
\begin{align}
  \Tilde{\beta} \approx - \frac{ 2 }{ \phi^2 \sin^2\chi },
\end{align}
which implies that $\beta_{\mr{iso}}$ becomes exponentially small.

\subsection{$\phis < v_\phi$ and $v_\phi = \mc{O}(1)$}

\begin{figure}[t]
  \centering
  \begin{subfigure}{0.32\textwidth}
    \centering
    \includegraphics[width=\textwidth]{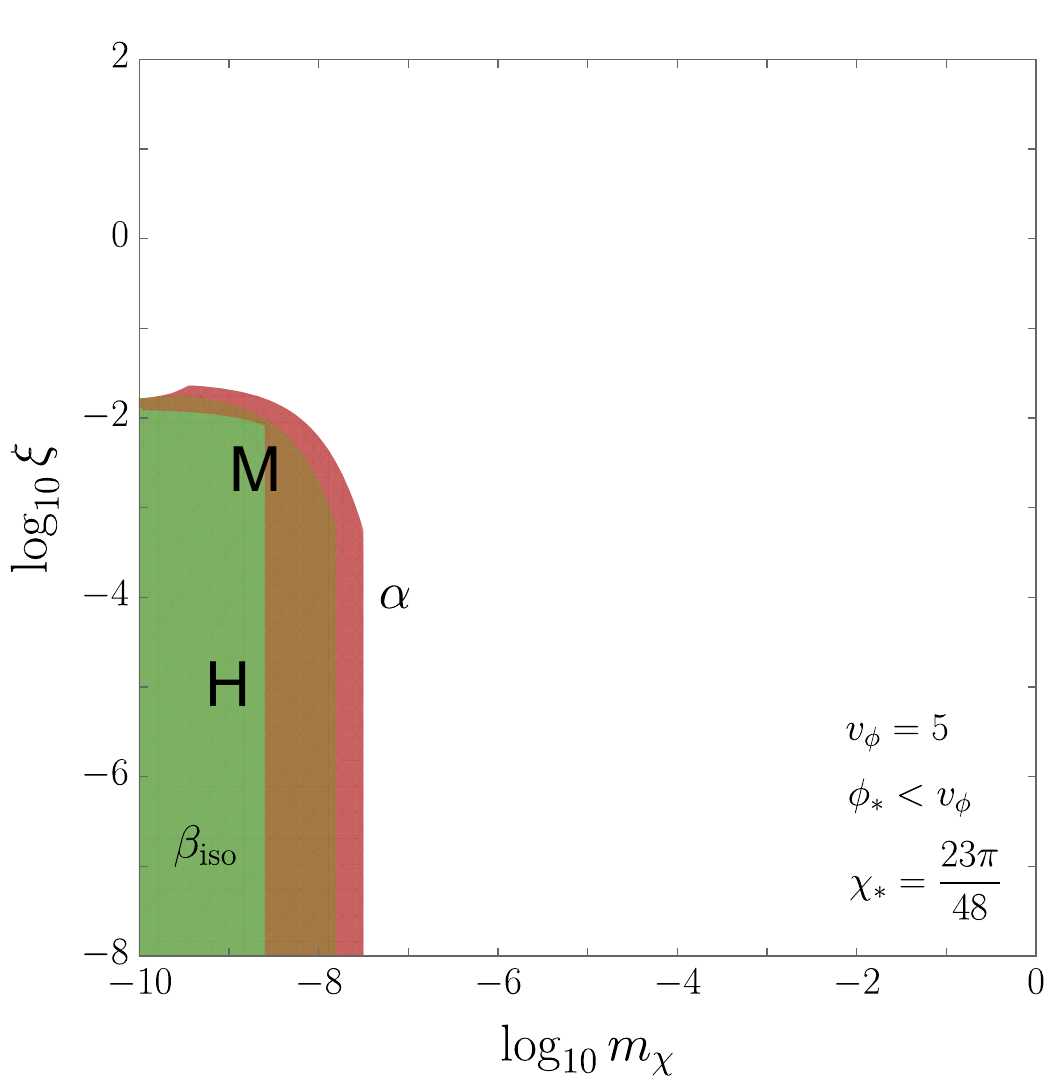}
    \caption{$\chis = 23 \pi/ 48$}
  \end{subfigure}\ 
  \begin{subfigure}{0.32\textwidth}
    \centering
    \includegraphics[width=\textwidth]{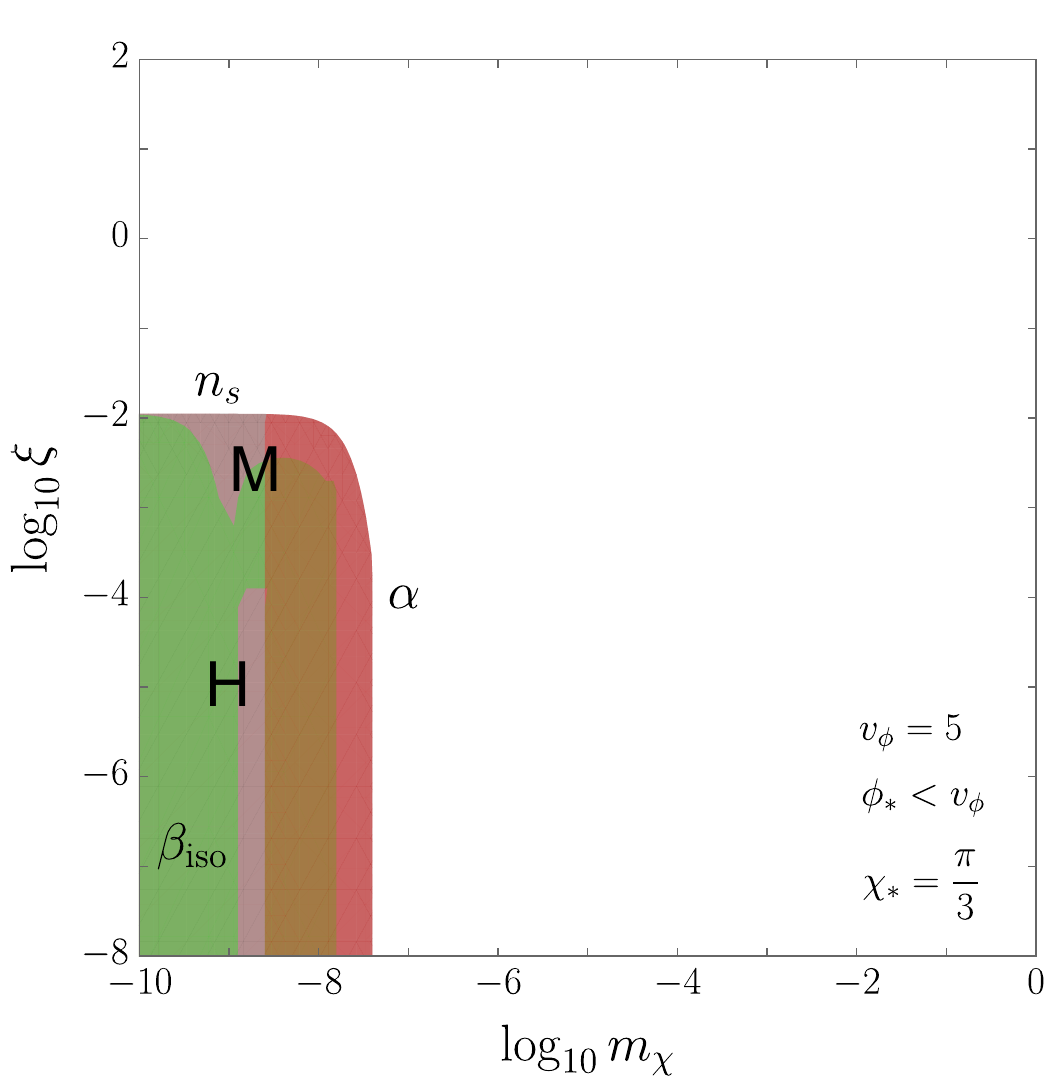}
    \caption{$\chis = \pi /3$}
  \end{subfigure}\ 
  \begin{subfigure}{0.32\textwidth}
    \centering
    \includegraphics[width=\textwidth]{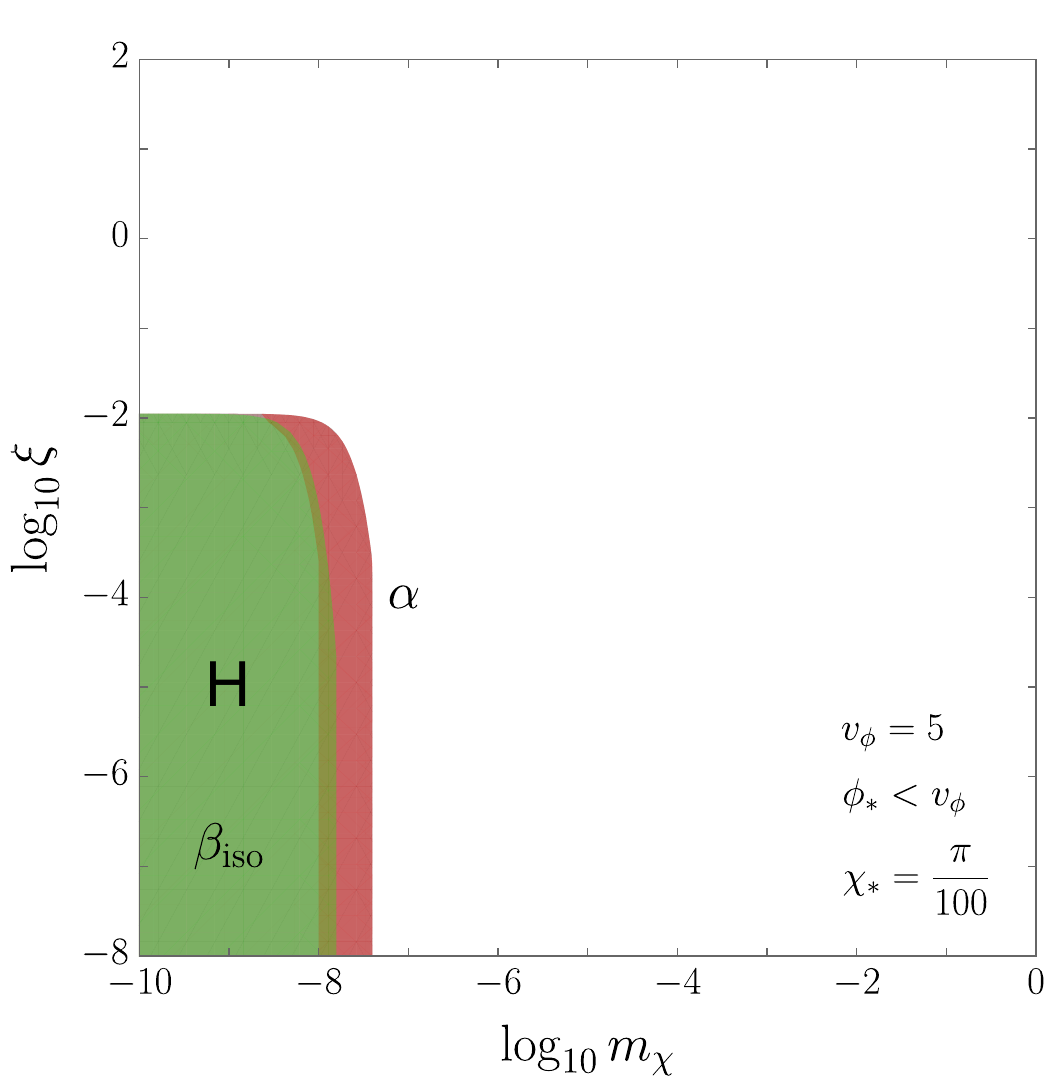}
    \caption{$\chis = \pi /100$}
  \end{subfigure}
  \caption{
    The parameter region in the $(m_\chi, \xi)$ plane for
    $\phis < v_\phi = 5$.
    The meanings of the colored regions are the same as in
    Figure~\ref{Figure:mchixiVEV20L}. 
    No experimentally allowed region is found. 
  }
  \label{Figure:mchixiVEV5}
  \bigskip
\end{figure}

As the final pattern of inflation in the present two-field model, we
consider the case with an intermediate symmetry-breaking scale.
Figure~\ref{Figure:mchixiVEV5} shows the experimentally allowed
parameter region for $\phis < v_\phi = 5$. 
It is found that most of the region is more severely constrained by the
condition $N=55$ compared to the large VEV case with $\phis < v_{\phi}$. 
A large value of $\xi$ is not allowed when $v_\phi$ is small, since the
potential must remain sufficiently flat for successful inflation. 
Furthermore, compared to the large VEV case, the region with large
$m_\chi$ is also excluded. 
This is because the inflaton potential becomes flatter for larger
$v_{\phi}$ and smaller $m_{\chi}$. 
For large $m_{\chi}$ and $\phis < v_\phi$, the power spectrum is
approximated by
\begin{align}
  \mc{P_{R}} \approx
  \frac{ \lambda^{3} v_{\phi}^{12} ( 1 + T_{\mc{RS}}^{2} ) }{ 1536 \pi^2 \phis^2
   [ \lambda^2 v_\phi^4 +4 ( \lambda v_\phi^{2} + m_{\chi}^{2} ) m_{\chi}^{2} \sin^2 \chis ]},
\end{align}
indicating that a small $v_\phi$ allows only a small $m_\chi$, as
required by the Planck normalization of $\mc{P_R}$.

The parameter region shown in Figure~\ref{Figure:mchixiVEV5}, especially
in the middle panel, is also constrained by inconsistency with the
observed value of $n_s$. An approximate expression for $n_s$ is obtained as
\begin{align}
  n_{s} \approx 1-\frac{8(1+\xi v_\phi^2)}{v_{\phi}^{2}}
  + \mc{O}( m_{\chi}^{2} ),
\end{align}
which shows that $n_s$ tends to take values smaller than the
observational bounds in this case~\cite{Barenboim:2013wra}. 
This behavior originates from the relatively large acceleration of the
inflaton field in this region. The tensor-to-scalar ratio is approximated as
\begin{align}
  r \approx \frac{128 \phis^{2}(1+\xi v_\phi^2)^2}{v_{\phi}^{4}}
  + \mc{O}(m_{\chi}^{2}),
\end{align}
and remains small throughout the region where the inflaton potential is
flat. The running $\alpha$ is approximately given by
Eq.~\eqref{Formula:Approalpha3}. 
In this case, the condition $\phi_{*}\ll1$ is realized, implying that
$\alpha$ is consistent with observations provided that $m_{\chi}$ is
sufficiently small.

Finally, we discuss the isocurvature fraction $\beta_\iso$. 
In this case, it is found to be large over most of the parameter space.
The approximation of $\tilde{\beta}$ is given by Eq.~\eqref{Formula:beta2}. 
In contrast to the large VEV case, we find that
$\tilde{\beta}\approx 0$ at the onset of inflation. 
This reflects the fact that the inflaton is initially almost insensitive
to the $\phi$ direction, leading to the approximation
$\Tilde{\beta}\approx 0$ holding for an extended period during inflation. 
Such behavior implies that a relatively small symmetry-breaking scale
under the condition $\phis < v_{\phi}$ results in a large
$\beta_\iso$, which is incompatible with observational constraints.

\newpage

{\small
\bibliographystyle{utphysm}
\bibliography{ref}
}

\end{document}